\documentclass[useAMS,usenatbib]{mn2e}

\usepackage{graphicx}
\usepackage{times}
\usepackage{amsbsy}
\usepackage{amssymb}
\usepackage{natbib}

\newcommand{\Mpc}{\mbox{ Mpc}}
\newcommand{\eV}{\mbox{ eV}}
\newcommand{\EeV}{\mbox{ EeV}}
\newcommand{\muG}{\ \mu\mbox{G}}

\newcommand{\enzo}{\it{\small ENZO}}

\newlength{\onewidth}
\newlength{\minipagewidth}
\newlength{\skywidth}
\newlength{\skycolumn}
\newlength{\threewidth}
\def \obs {61}
\def \simulation {agn}
\begin{document}

\pubyear{2016}
\volume{e-print}
\label{firstpage}
\pagerange{\pageref{firstpage}--\pageref{lastpage}}
 
\title[Propagation of UHECRs in EGMFs]{Propagation of Ultra High Energy Cosmic Rays in Extragalactic Magnetic Fields: A view from cosmological simulations}
\author[S. Hackstein et al.]{S. Hackstein$^{1}$\thanks{
 E-mail: shackste@physnet.uni-hamburg.de}, F. Vazza$^{1}$, M. Br\"{u}ggen$^{1}$, G. Sigl$^{2}$, A. Dundovic$^{2}$\\
$^{1}$ Hamburger Sternwarte, Gojenbergsweg 112, 21029 Hamburg, Germany\\
$^{2}$ Universit\"at Hamburg, II. Institut f\"ur Theoretische Physik, Hamburg, Germany}

\date{Accepted 2016 July 27. Received 2016 July 20; in original form 2016 June 13}
\maketitle

\begin{abstract}
	We use the CRPropa code to simulate the propagation of ultra high energy cosmic rays (with energy $\geq 10^{18} \eV$  and pure proton composition) through extragalactic magnetic fields that have been simulated with the cosmological ENZO code. 
	We test both primordial and astrophysical magnetogenesis scenarios in order to investigate the impact of different magnetic field strengths in clusters, filaments and voids  on the deflection of cosmic rays propagating across cosmological distances.
	We also study the effect of different source distributions of cosmic rays around simulated Milky-Way like observers.
	Our analysis shows that  the arrival spectra and anisotropy of events are rather insensitive to the distribution of extragalactic magnetic fields, while they are more affected by the clustering of sources within a $\sim 50$ Mpc distance to observers.
	Finally, we find that in order to reproduce the observed degree of isotropy of cosmic rays at $\sim $ EeV energies, the average magnetic fields in cosmic voids must be  $\sim 0.1 \rm \ nG$, providing limits on the strength of primordial seed fields.
\end{abstract}


\section{Introduction}
\label{sec:intro}
	Cosmic Rays (CRs) mostly consist of charged nuclei that travel through space and may enter the atmosphere of Earth, producing  a cascade of particles \citep[e. g.][]{CRreview1,CRreview2}.
	They are deflected by the Lorentz  force of cosmic magnetic fields (CMF), and this complicates the search for their origin as the events do not directly point back to their sources.
	At low energies, $< 10^{17} \eV$, the galactic magnetic field (GMF) of the Milky Way \citep[$\sim3-6\muG$, e. g.][]{MW_field} is strong enough to confine the CRs within the galaxy.
	The most plausible sources of these galactic CRs are shocks in the remnants of supernovae which diffusively accelerate particles \citep[e. g.][]{Blasi}.
\\ \\
	At energies higher than $10^{17} \eV$, the gyro-radius of the CRs in typical GMFs \citep[$\sim 5-15\muG$, e. g.][]{galactic_fields} becomes so large that they cannot be confined to galaxies any more. Therefore, energies between $10^{17}$ and $10^{19} \eV$ are expected to mark the transition from galactic to extragalactic sources \citep[e. g.][]{Aloisio:2012ba}.
	The sources of ultra high energy cosmic rays (UHECRs) have not yet been identified.
	Finding these sources, as well as  finding the composition of UHECRs at the highest energies, represent some of the most important challenges in this field of research.\\ \\ \\
	When they travel through space, UHECRs interact with ambient photon fields, produce secondary CRs\footnote{e. g. photons, neutrinos, muons etc., which are not considered here.} in pair production processes and thereby lose parts of their energy \citep[e. g.][]{Eloss}.
	At energies above the GZK-threshold $E_{\rm GZK} \approx 4\cdot10^{19} \eV$, UHECRs are energetic enough to produce pions via $\Delta$-resonance \citep{Greisen,Zatsepin_Kuzmin_1966,confirm_GZK}.
	This causes UHECRs to lose energy over rather short distances, thereby restricting the maximum distance they can travel  to $\sim 100 \Mpc$, the so called GZK-Horizon. \citep[e. g.][]{GZK_horizon1,GZK_horizon2}
\\ \\
	The study of UHECRs propagating in extragalactic magnetic fields (EGMFs) is made even more complicated by the fact that the present strength, topology as well as the origin of EGMFs are unknown \citep[e. g.][]{Widrow2011}. 
	While the shape, strength and structure of magnetic fields of galaxies and galaxy clusters have been measured to some extent \citep[e. g.][]{Feretti2012,Bonafede2013}, the distribution of  magnetic fields in cluster outskirts, filaments and voids remains largely unknown. 
	In particular, outside of galaxy clusters the predicted strength of extragalactic magnetic fields diverge when different models of magnetic field seeding are considered, i.e. seeding from primordial fields produced during inflation or baryogenesis \citep[e. g.][]{Widrow2011,2015arXiv150402311S}
	versus seeding from magnetised galactic winds  \citep[e. g.][]{Bertone2006,Donnert2008} and active galactic nuclei (AGNs) \citep[][]{xu09}.
	Recent results from the PLANCK collaboration have put an upper limit on the strength of a primordial field at the surface of the last scattering of the order of $B_0 \leq 0.55-5.6 \rm nG$ \citep[][]{PLANCK2015}.
	The additional analysis of non-Gaussianity in the Cosmic Microwave Background have further suggested slightly lower upper limits, in the range $B_0 \leq 0.05-0.6 \rm nG$ \citep[][]{2014PhRvD..89d3523T}.
	These limits translate into a slightly lower limit on the present-day magnetisation of voids in absence of other sources of magnetisation, $B_{\rm void} \approx B_0 (\rho_{\rm v}/\langle \rho \rangle)^{2/3}$ (where $\rho_{\rm v}$ is the gas density in voids and $\langle \rho \rangle $ is the average gas density of the Universe).
	 On the other hand, a lower limit of the order of $\geq 10^{-16} \rm G$ for the magnetisation of voids has been suggested by \citet[][]{NeronovVovk2010} based on the absence of GeV secondary emission detected around TeV blazar sources.
	However, it shall be noticed that the validity of these lower limits is still subject to debate \citep[see discussion in ][]{Broderick:2011av}.
	 While future radio observations will have the chance  to probe the magnetisation of cluster outskirts and filaments \citep[e. g.][]{Vazza2015}, the study of UHECRs has the unique potential to probe the magnetisation level of voids in the local Universe (and hence also the amplitude of primordial magnetic fields). 
	Moreover, the possibility of limiting extragalactic magnetic fields outside clusters would also be crucial to assess the viability of axion-like particles as a candidate for dark matter because in the presence of significant magnetic fields they produce detectable effects on the spectrum of TeV sources \citep[e. g.][]{2012PhRvD..86g5024H}.
\\
	In this paper, we study whether realistic models of EGMFs can significantly affect the propagation of UHECRs from nearby sources.
\\ \\
	One of the most promising possible sources is Centaurus A, which contains the most nearby radio-loud active galactic nucleus. 
	The Pierre Auger Collaboration has tentatively reported the detection of an excess of UHECR events around Cen A\citep{Auger2010}. However, the statistical significance of this detection is uncertain.  
	In this paper we also investigate possible interpretations of this tentative detection. 
	Furthermore, we present a technique to compare the experimentally observed separation of events with other sets of possible sources of UHECRs.
\\ \\
	This paper is organized as follows:
	In Sec. \ref{sec:methods} we briefly discuss  the simulation methods we used in this work and present the magnetic field models we investigated.
	The results of this investigation are presented in Sec. \ref{sec:results}, where we analyse the energy spectrum, the angular power spectrum and the angular separation of observed UHECR events.
	Our conclusions are given in Sec. \ref{sec:conclusions}.

\section{Methods}
\label{sec:methods}
\subsection{Magnetic field modelling}
\label{subsec:ENZO}

	The magneto-hydrodynamics (MHD) simulations analysed in this paper have been produced with the cosmological grid code {\enzo}. {\enzo} is based on a particle-mesh N-body method 
 to follow the dynamics of the Dark Matter 
 and a variety of shock-capturing Riemann solvers to evolve the gas component  \citep{ENZO}.
	We solved the MHD equations employing the method by  \citet{Dedner2002}, which uses hyperbolic divergence cleaning to keep $\nabla \cdot \boldsymbol{B}$ as small as possible, and the 
 Piecewise Linear Method as a reconstruction method for the fluxes at cell interfaces, which are evolved using the local Lax-Friedrichs Riemann solver \citep{KurganovTadmor2000}, with time integration using the total variation diminishing 
 second order Runge-Kutta  scheme \citep{ShuOsher1988}.
	A subset of our simulations also made use of the recent porting of the Dedner algorithm onto CUDA \citep{WangAbel2010}, which runs $\sim 4$ times faster on Graphics Processing Units (GPU), compared to the performance on CPUs.
	This suite of  runs belongs to a larger suite of MHD cosmological simulations ("CHRONOS++"), designed to investigate the origin of extragalactic magnetic fields \citep[][]{Vazza2014,Vazza2015}.
\\ \\
\begin{figure}
\centering
\includegraphics[width=\onewidth]{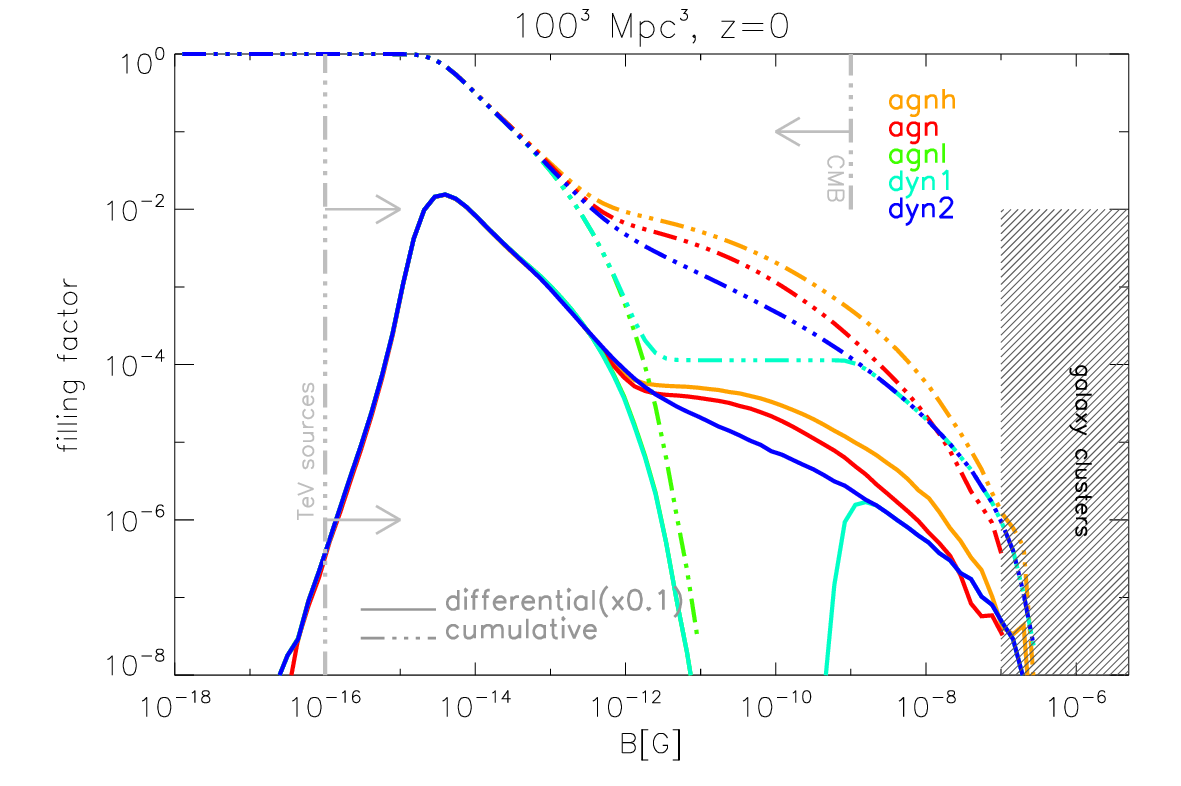}
\caption{
Cumulative and differential (rescaled by $0.1$ for clarity) volume filling factor at $z=0$ for the simulated magnetic fields. We additionally show for comparison the range of values expected in clusters, the possible lower limits from the magnetisations of voids \citep{NeronovVovk2010}, and the range of upper limits allowed by PLANCK \citep{Planck2013}.}
\label{fig:Bhisto}	
\end{figure}
\begin{table*}
	\centering
	\tabcolsep 5pt 
	\begin{tabular}{c|c|c|c|c}
		 $L_{\rm box}$ & $N_{\rm grid}$ & $\Delta x$ & physics & mnemonic \\\hline
		 [$\rm Mpc$]  &              & [$\rm kpc$]  &   & \\  \hline
		 100 & $512^3$ & $195$ & non-radiative, no magnetic field &  {\it b0} \\ 
		 100 & $512^3$ & $195$ & cooling, star formation, $B_0=10^{-14} \rm G$ + dynamo($n/n_{\rm cr} \geq 50 $) &  {\it dyn1} \\ 
		 100 & $512^3$ &$195$ &  cooling, star formation, $B_0=10^{-14} \rm G$ + dynamo($n/n_{\rm cr} \geq 1 $) &  {\it dyn2} \\ 
		 100 & $512^3$ &  $195$ & cooling, star formation, $B_0=10^{-14} \rm G$ + AGN($10^{57} \rm erg$) &  {\it agnl} \\ 
		 100 & $512^3$ &  $195$ &  cooling, star formation, $B_0=10^{-14} \rm G$ + AGN($10^{58} \rm erg$) &  {\it agn} \\ 
		 100 & $512^3$ &  $195$ &  cooling, star formation, $B_0=10^{-14} \rm G$ + AGN($10^{59} \rm erg$) &  {\it agnh} \\ \hline
		 100 & $512^3$ &  $195$ &  cooling, star formation, $B_0=10^{-13}-10^{-8} \rm G$ + AGN($10^{57} \rm erg$) &  {\it agnl+1...+6}
	\end{tabular}
	\caption{List of the simulations run for this project. First column: box length of the simulated volume;  second column: number of grid cells in the initial conditions;  third column: spatial resolution; fourth column: physical module; fifth column: name of the simulation. The last row refers to the ensemble of 5 additional runs with increased primordial seed fields, discussed in Sec. \ref{subsec:plusB} .}
	\label{tab:sim}
\end{table*}
	Here we study the propagation of UHECRs through a $(100 \rm Mpc)^3$ comoving volume, which has been simulated from $z=30$ to $z=0$ on a fixed grid of $512^3$ cells and using $512^3$ dark matter particles.	
	In addition to gravity and magneto-hydrodynamics,  our runs included metallicity-dependent equilibrium gas cooling and star formation  and feedback \citep{CenOstriker1992}.
	In order to bracket the realistic distribution of magnetic fields in regions yet to be observed we simulated the seeding of extragalactic magnetic fields in a variety of scenarios.
         In most of our runs we initialised the magnetic field at $z=30$ to the uniform value of $B_0=10^{-14} \rm G$ along each coordinate axis. {\footnote{{ This rather low value ensures that at $z=0$ the magnetisation levels of simulated cosmic voids does not fall below the lower limits derived by \citet[e.g.][]{NeronovVovk2010}, as discussed below in Fig.\ref{fig:Bhisto}. A complete survey of models with initial fields from $10^{-14} \rm G$  to $10^{-8} \rm G$ is given in Sec.~\ref{subsec:plusB}. }}}
	In absence of other seeding processes and at this rather coarse resolution, the comoving magnetic field is found to grow by compression, up to small values, $\sim 10^{-11} \rm G$, largely inconsistent with the $\sim\rm\mu G$ values of galaxy clusters.
	This is expected, because efficient dynamo amplification within structures can develop only if the gas flow is  turbulent enough, which requires much larger resolution \citep[e. g.][]{Ryu2008, Beresnyak2015}. 
\\
	To overcome the limitations of resolution, similar to \citet{Vazza2015}, we renormalized the field strength of each cell in post-processing so that the magnetic field energy is $1\%$ of the thermal gas energy  as in a saturated small-scale dynamo scenario.
	However, while dense structures such as galaxy clusters are expected to host small-scale dynamo \citep[e. g.][]{Beresnyak2015}, the situation is less clear in cluster outskirts and filaments \citep[e. g.][]{Vazza2014}.  
	In order to bracket the uncertainties of where the small-scale dynamo process starts operating, we produced two versions of this run, by renormalising the magnetic energy wherever the gas density is $\geq \rho_{\rm cr,g}$ (where $\rho_{\rm cr,g}$ is the critical gas density) or only limited to where the gas density is $\geq 50 \rho_{\rm cr,g}$. 
\\
	The first case represents a scenario in which small-scale dynamo operates everywhere in the overdense cosmic web ({\it dyn2} model), while the second case represents a scenario in which the dynamo amplification operates only within virialised halos ({\it dyn1} model).
	The two above models investigate a fully "primordial" scenario for extragalactic magnetic fields, while in a second set of runs we investigate an "astrophysical" scenario for the magnetisation of large-scale structures. 
	In this case, we allowed for the impulsive thermal and magnetic feedback within those halos, where the physical gas density exceeded the critical value of $10^{-2} \rm cm^{-3}$, which marks the onset of catastrophic gas cooling. 
\\
	To bracket the range of AGN energies, we varied the total energy of each AGN feedback event from $10^{57} \rm erg$ ({\it agnl}) via $10^{58} \rm erg$ ({\it agn}) to $10^{59} \rm erg$ ({\it agnh}). The magnetic energy is always assumed to be a fixed $1\%$ of the injected thermal energy. 
	While the thermal energy is released as  a couple of overpressurised outflows at random opposite directions from the halo centre, the feedback magnetic energy is released as dipoles around the halo centre. 
	With similar runs (without magnetic fields) in \citet{Vazza2016} we showed that a fixed $10^{59} \rm erg$ energy for events can roughly reproduce the $(M,T)$ observed scaling relation of galaxy clusters and groups as well as broadly reproduce the  bimodality of radial profiles in clusters.
	Finally, we simulated the same volume without magnetic fields ({\it b0}) as a control run. 
	A schematic view of the simulations used in this work is given in Tab. \ref{tab:sim}.
\\ \\
	In Fig. \ref{fig:Bhisto} we show the differential and cumulative volume filling factor of magnetic fields, i. e. the part of the volume filled with magnetic field of strength equal or above the given value, for various models.
	In order to give a sense of the available observational constraints in the same figure we also show the possible lower limit on the {\it present} magnetisation of voids \citep[][]{NeronovVovk2010}, the range of upper limits from PLANCK \citep{Planck2013} and the typical range of cluster magnetic fields. 
	Due to the chosen small initial magnetic field seed  most of the cosmic volume has a very low magnetisation level in all simulated scenarios, i.e. $\geq 90\%$ of the volume has $B \leq 10^{-12} \rm G$.
	All primordial scenarios {\it dyn1} and {\it dyn2} (after rescaling for the unresolved small-scale dynamo effect) and the astrophysical scenarios {\it agn} and {\it agnh} achieve $\sim 0.1 \rm \ \mu  G$ levels in galaxy clusters and groups with slightly different filling factors depending on the seeding recipe. 
	The prediction from the different models are maximally different for $\rho \sim (1-50) \rho_{\rm cr,g}$, i. e. the over-density regime of filaments and cluster outskirts. 
	The fields can be of the order of $\sim \rm nG$ there in the {\it agn} and {\it agnh} scenario, $\sim 0.1 \rm ~nG$ in the {\it dyn2} scenario, or as low as $\sim 0.001 \rm ~nG$ in the other cases. 
\\
	All our investigated scenarios ensure a range of magnetic field values which is within the bounds of observations and all runs (with the exception of the {\it agnl} run) also reproduce typically observed magnetic fields in galaxy clusters{\footnote{We remark that in our $100^3 \rm ~Mpc^3$ only one galaxy cluster with a size comparable to the Coma cluster is formed, hence we compare with the magnetic fields of lower mass systems here}}.
	We notice that in all our models the minimum magnetisation level is much smaller than what is assumed in other works studying UHECRs with simulations, e. g. \citet{Sigl2004}, where the magnetic field is $\sim 1-10 \rm\ nG$ in voids. 
	This is because in the latter works a fixed {\it global} rescaling of the simulated magnetic fields was performed in order to match the values of clusters, while our renormalisation here scales with gas energy. 
\\
	In a second suite of runs, we globally rescaled the {\it agnl} model to reach filling factors similar to earlier models and compare with earlier results. This allows us to probe the influence of magnetic fields in filaments and cluster outskirts on UHECR observables, as well as the influence of the field strength in voids.
	In Sec. \ref{subsec:plusB} we will present additional results, where in post-processing we rescaled the strength of the magnetic field everywhere in the simulations in steps of factors of 10, from 10 to $10^6$.
	This process is entirely equivalent to increasing  the strength of the primordial seed field from $B_0=10^{-14} ~\rm G$ to $B_0=10^{-8} ~\rm G$ in order to assess the effect of the magnetisation in voids on the propagation of UHECRs. 
	We perform this rescaling only in our {\it agnl }case, which is essentially unaffected by the impact of AGN, and is therefore fully equivalent to a standard MHD run where the magnetic fields are only due to cosmological seeding.
	However, we notice already that the effects are significant only for a range of primordial magnetic field strengths ($\geq 10^{-10} \rm ~ G$) which are in tension with the latest PLANCK constraints or can already be rejected based on them. 

\subsection{CR Propagation}
\label{subsec:CRPropa}
\begin{figure}
\centering
\includegraphics[width=\onewidth]{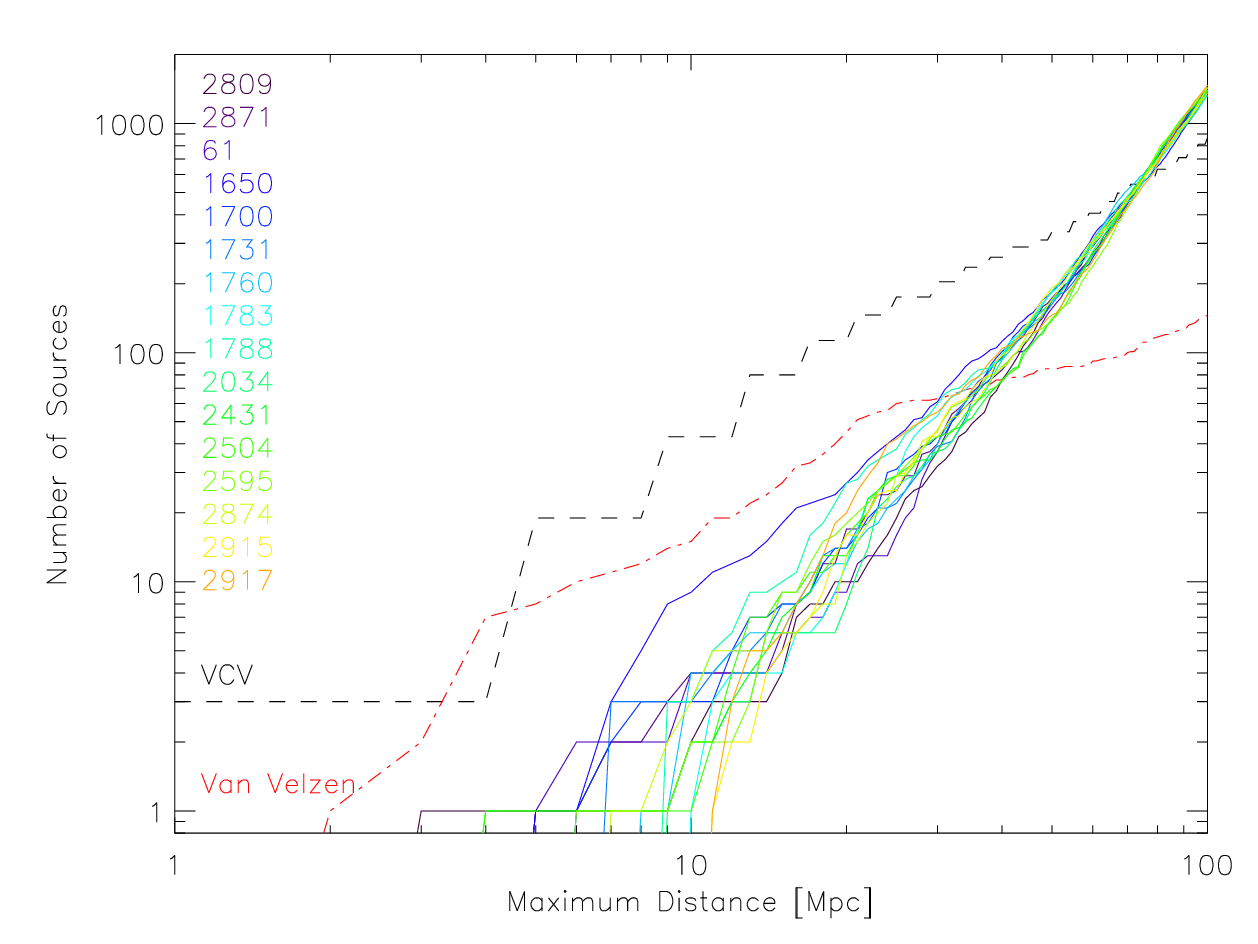}
\caption{Number of sources within given distance from the observer for the 16 different observers in the MHD simulations and number of AGN  \citep{VCV} or radio sources \citep{Velzen} from catalogues of the real universe.}
\label{fig:source_histo}
\end{figure}
	In this work we made use of CRPropa 3.0\footnote{http://github.com/CRPropa/CRPropa3}
	\citep{CRPropa2006,CRPropa2013,CRPropa2016}
	which is a publicly available code to simulate the propagation of UHECRs in 3D CMF-simulations.
	CRPropa injects a particle into a given CMF simulation with some initial momentum and then computes its propagation and energy losses.
	The path of propagation is calculated step by step by integrating the Lorentz equations and computing particle interactions. 
	Pair production is approximated as a continuous energy loss and
	pion production by using SOPHIA\footnote{Simulations Of Photo Hadronic Interactions in Astrophysics},
	which is a Monte Carlo event generator designed to study this process \citep{Mucke1999}.
	In the propagation of UHECRs, we assumed the computational volume to be periodic.
	Any time a cosmic ray hits an observer sphere, an event is recorded.
	The final output of CRPropa is a set of positions, momenta and energies of the initial and final state
	of all particles that hit the observer sphere as well as the distance they have traveled.
\\ \\
	In the CRPropa runs presented here we used observer spheres of radius $0.8 \Mpc$, which is roughly the distance to our closest neighbouring galaxy M31. 
	In each run, we injected $10^7$ protons with energies from $10^{18} \eV$ to $10^{21} \eV$ with a spectral index of $-2$, resulting in $\sim 7\cdot 10^4$ observed events for every observer, enough to converge the energy spectra.
	This set of parameters allows for a physically relevant and sufficiently large set of data in feasible computation time.
\\
	These runs were repeated for 16 different observer positions in the 5 different magnetic field models and in absence of magnetic fields.
	The roughly 300 sources of CRs in the simulated volume were selected to be at the centre of all halos with a total mass threshold of $10^{11} M_{\odot}$, identified at $z=0$. 
	The observer positions were drawn from isolated halos with masses of the order of the Milky Way.
	Fig. \ref{fig:source_histo} shows a comparison of the source distance distribution of all observers with those of AGN and of radio sources from catalogues of the real universe \citep{VCV,Velzen}. 
	The number of sources within 100 Mpc of the simulated observers is comparable to the real number of AGN or radio sources.	They are therefore a suitable choice to probe the scenario where these objects are considered as the sources of UHECRs.
\\
	In total, the combination of the 6 magnetic field models and of 16 observers in each model gives a final amount of 96 runs with CRpropa3, where we followed the propagation of about $10^{9}$ protons in order to collect $\sim 7 \cdot 10^{6}$ events on the observers' spheres. 
\subsection{Finite Observer Effect}
\label{subsec:geometriceffect}
\begin{figure}
\centering
\begin{minipage}[b]{\minipagewidth}
	\includegraphics[width=\onewidth]{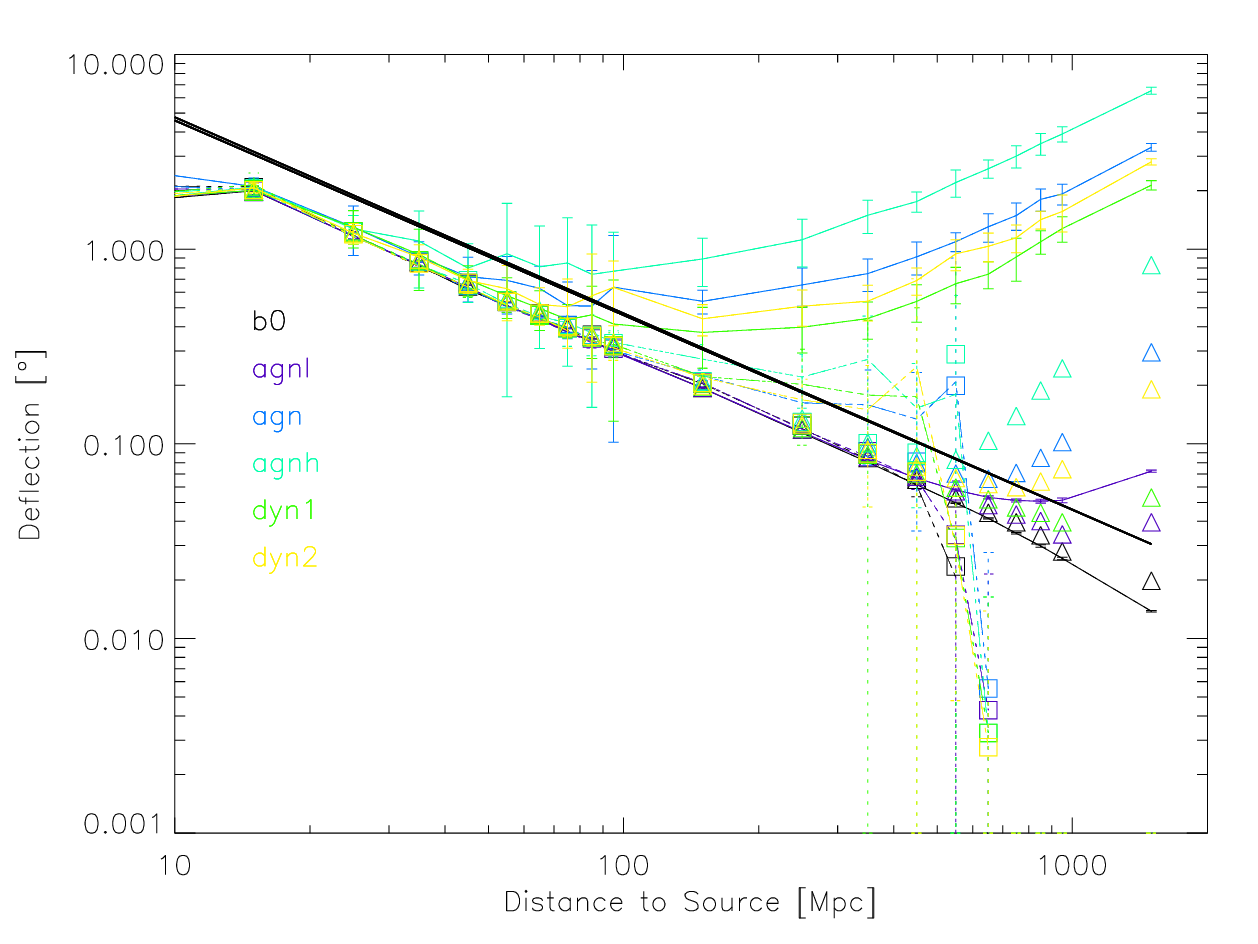}
	\caption{Angle between computed arrival momentum and line of sight between observer
center and source, see Fig. \ref{fig:geometricdeflection}, as a function of the distance to the source of the UHECR. The solid/dash-dotted line is the average of all events with energies $E > 1, \ 58$ EeV, respectively. The triangular/quadratic symbols show the median. The error bars show the $1\sigma$ standard deviation. The black line is the prediction of the maximum artificial deflection angle below which deflection angles cannot be trusted within this approach.}
	\label{fig:deflection_angle}
\end{minipage}
\end{figure}
	On typical length scales for UHECR propagation, the size of the Earth as an observer vanishes, so ideally in a simulation a point-like observer should be assumed.
	However, such a simulation is very inefficient as only very few events reach the observer, and the computation is dominated by the propagation of particles that do not reach the observer \citep{finite_observer}.
	On the other hand, a finite size of the observer can introduce spurious effects in the measurements of the anisotropy of UHECRs, and must be treated with care (see below).
\\ \\
	The deflection angle $\Theta$ is the angular distance between the arrival direction of a particle and the vector that points from observer to its source and is computed via the scalar product.
\begin{equation}
	\Theta = \arctan \left( \frac{\boldsymbol{p}_1\cdot\boldsymbol{p}_2}{|\boldsymbol{p}_1||\boldsymbol{p}_2|} \right) . 
\end{equation}
	In Fig. \ref{fig:deflection_angle} the average deflection angle of UHECRs as a function of the distance to their sources is shown. 
\begin{figure}
\centering
	\includegraphics[width=\onewidth]{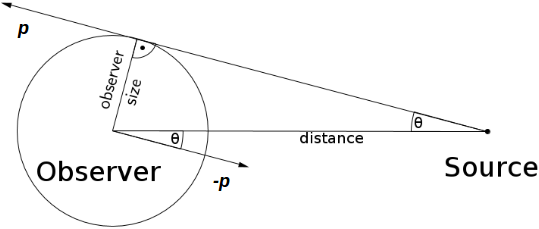}
	\caption{Schematic view on the case of maximum artificial deflection}
	\label{fig:geometricdeflection}
\end{figure}
	In the absence of magnetic fields a particle from a close-by source that tangentially hits the sphere of a finite observer is still recorded as observed (cf. Fig. \ref{fig:geometricdeflection}).
	The artificial geometrical deflection due to the observer's size (i. e. $\Theta_{art} = \arcsin\left(\frac{{\rm observer \ size}}{{\rm distance}}\right)$) dominates over the Lorentzian deflection for close-by sources.
	We emphasize that the usual uncertainty on arrival directions in experiments is of the order of $\sim 1 ^\circ$.
	From Fig. \ref{fig:deflection_angle} we see that the actual deflection of UHECRs that originate within 100 Mpc is negligible.
	On the other hand, the artificial deflection of the finite observer effect for particles from sources within 30 Mpc exceeds the uncertainty of experiments.
	The closest sources cause the largest number of events, hence the geometrical deflection will strongly affect the simulated observables, e. g. by smearing out the actual anisotropy signal.
	Therefore, in the interpretation of the following statistics the finite observer effect has to be treated with care. 
\section{Results}
\label{sec:results}
\begin{figure}
\centering
	\includegraphics[width=\skycolumn]{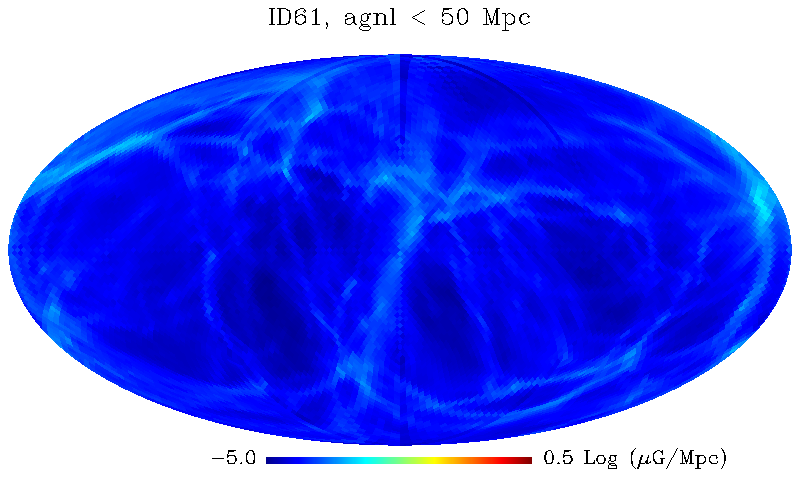}
	\includegraphics[width=\skycolumn]{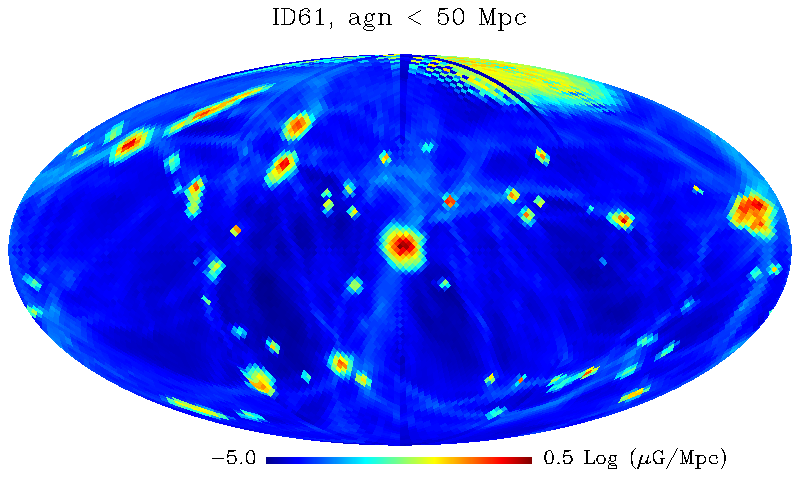}
	\includegraphics[width=\skycolumn]{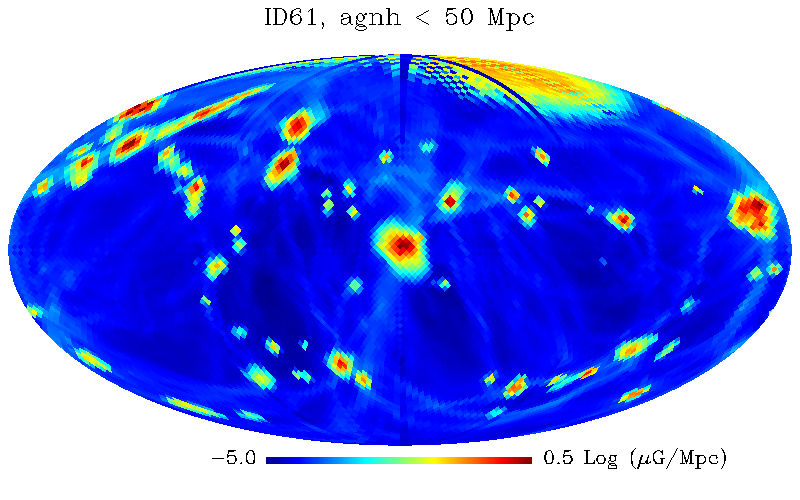}
	\includegraphics[width=\skycolumn]{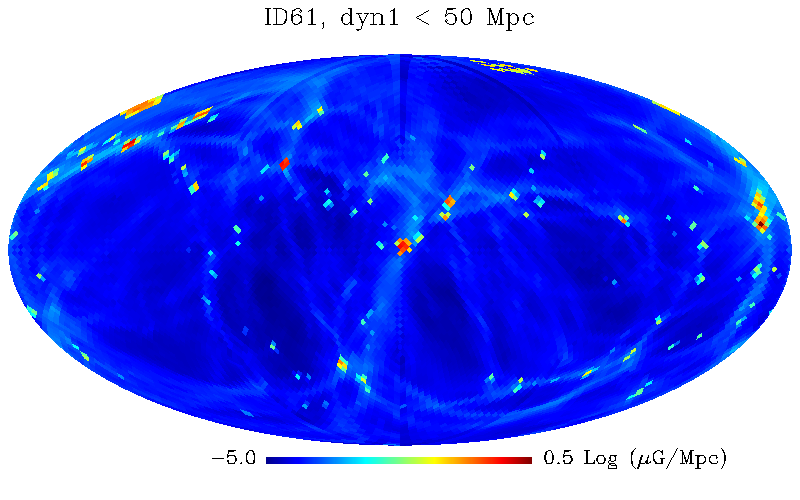}
	\includegraphics[width=\skycolumn]{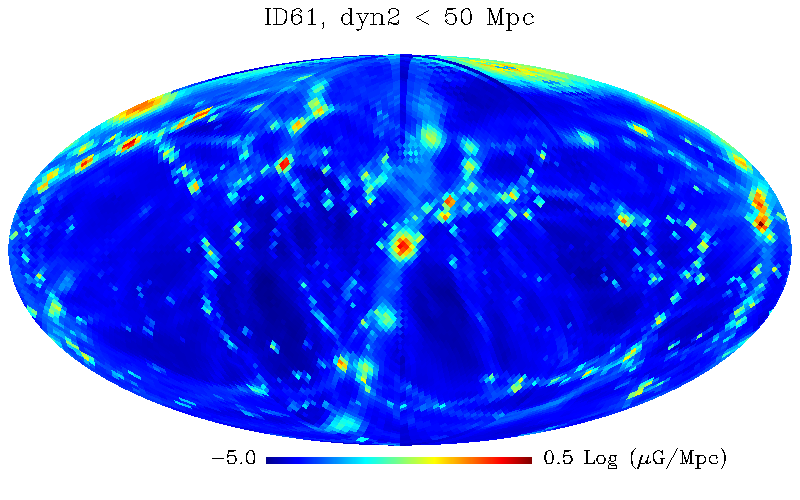}
	\caption{Full-sky map of magnetic field strengths within $50 \ \rm Mpc$ distance to the observer ID61 weighted by their distance in the 5 different models presented in Tab. \ref{tab:sim} (top to bottom: {\it agnl}, {\it agn}, {\it agnh}, {\it dyn1}, {\it dyn2}) .} 
	\label{fig:fullsky_Bmap}
\end{figure}
\begin{figure*}
	\includegraphics[width=\skywidth]{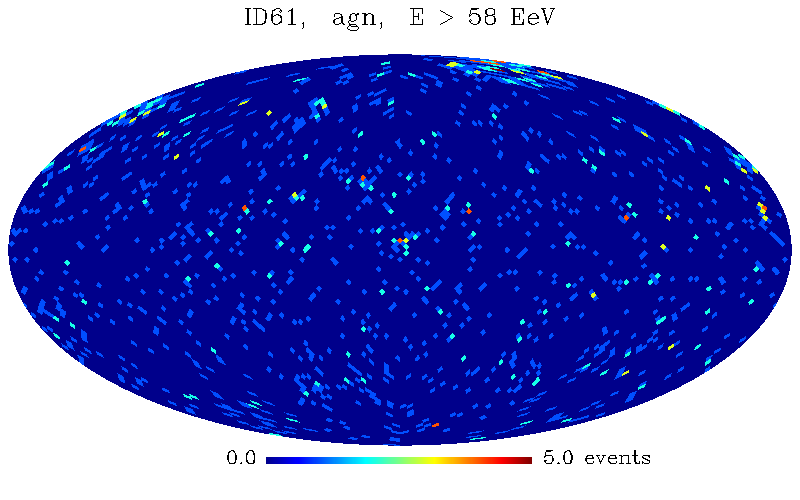}
	\includegraphics[width=\skywidth]{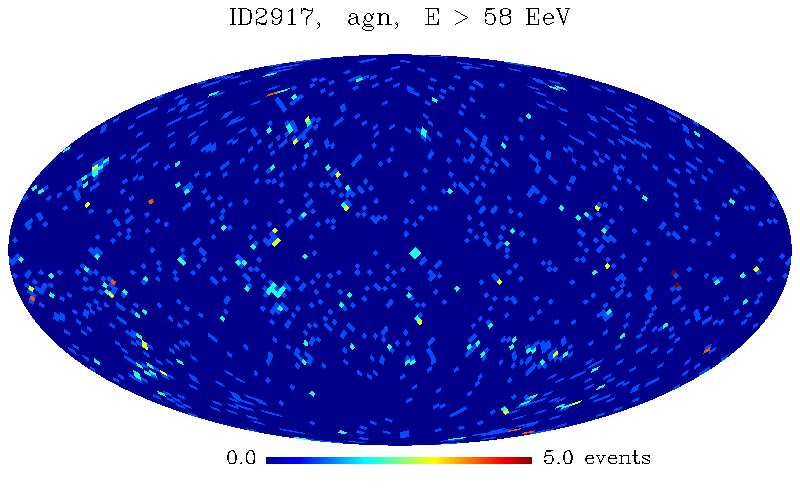}
	\caption{Full-sky maps of arrival direction of observed UHECR events with minimum energy $E > 58$ EeV for observer ID61 and ID2917 in the {\it \simulation} scenario.}
	\label{fig:fullsky_events}
	\label{fig:fullsky2917}
\end{figure*}
\begin{figure*}
	\includegraphics[width=\skywidth]{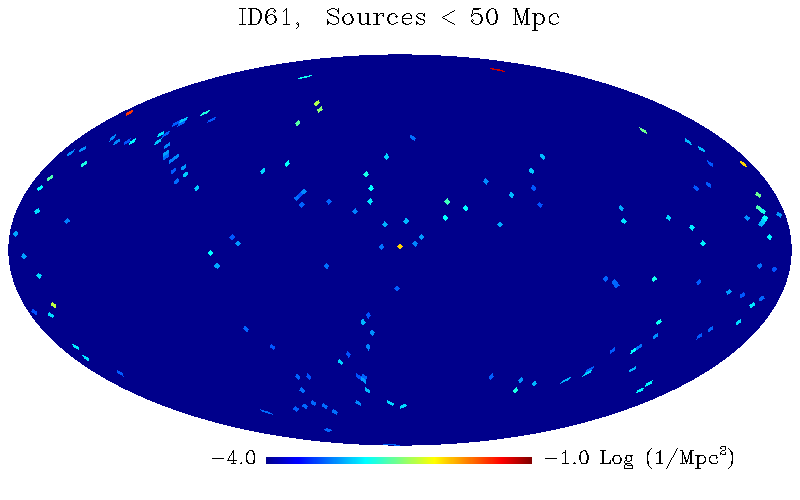}
	\includegraphics[width=\skywidth]{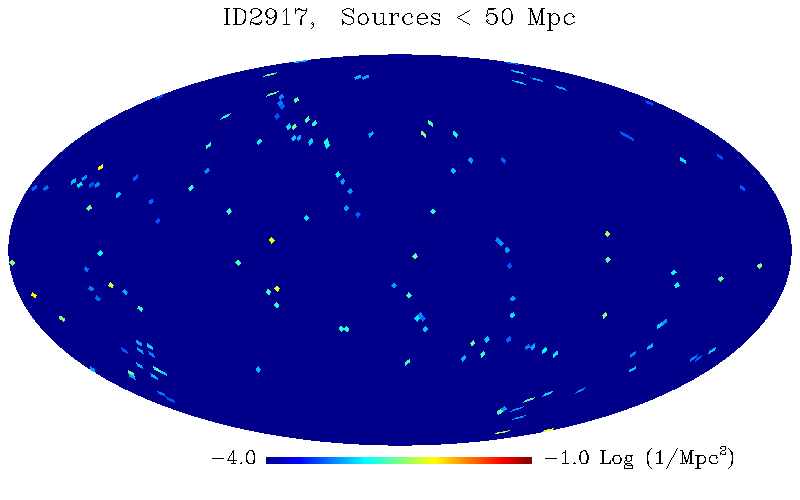}
	\caption{Full-sky maps of sources within 50 Mpc distance to the observer weighted by their inverse distance squared for observers ID61 and ID2917.}
	\label{fig:fullsky_sources}
\end{figure*}

{
 	 In the following presentation of our results we  will show that the distribution of sources and magnetised structures around the observer is crucial to  determine several of the observed properties of UHECRs. 
	Therefore, we begin by giving the visual impression of the projected distribution of magnetic fields obtained in our different resimulations. In this case, we give the example of the extragalactic magnetic fields
around observer ID61, which will be also used later in detail. 
	The magnetic field maps in Fig. \ref{fig:fullsky_Bmap} show the integral over the $|\vec{B}|$-values along the line of sight from ID61, normalised to the total distance of each casted ray. 
	These maps well illustrate the differences between the magnetic field models considered here.
	Fig. \ref{fig:fullsky_events} shows the arrival direction, i. e. the normalised negative momentum vector, of all events with $E > 58 \EeV$ observed by ID61 and ID2917.
	The maps  in Fig. \ref{fig:fullsky_sources}  show the inverse square of the distance of sources around these observers.
\\
	Consistent with the volume filling factors presented in Sec.\ref{subsec:ENZO}, we can see that in both the models with seeding of magnetic fields from AGN and where a dynamo amplification is assumed the magnetic field strength inside galaxy clusters and groups is similar. 
	However, outside of the viral radius of these structures the differences between models are more marked, and the action of AGN feedback is capable of magnetising the cosmic volume to a larger radius from the centre of structures. 
	On the other hand, if no dynamo amplification or AGN seeding is considered (top panel) the magnetic fields in clusters are too small compared to observations.
}

\subsection{Energy Spectrum}
\label{subsec:energyspectrum}
\begin{figure}
\centering
	\includegraphics[width=\onewidth]{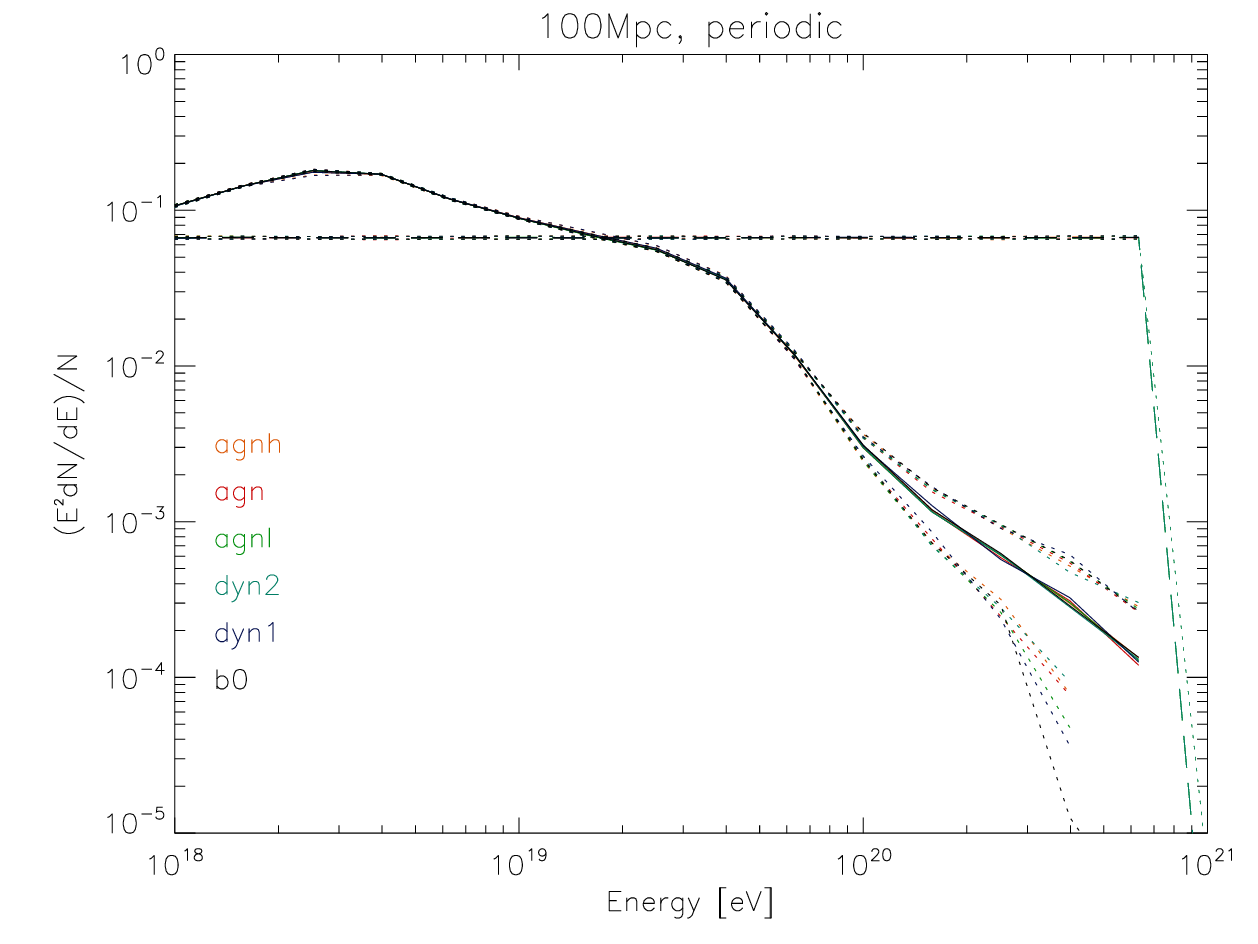}
	\caption{Energy spectrum of observed UHECR events averaged over all 16 observers in the different magnetic field models. The dashed line shows the injected, the solid line the predicted spectrum at the observer. The dotted lines show the $1\sigma$ standard deviation.}
	\label{fig:energyspectrum61}
	\label{fig:energyspectrum}
\hspace{1.5cm}
\end{figure}
	The energy spectrum of the observed UHECR events is shown in Fig. \ref{fig:energyspectrum61} for the six different magnetic field models averaged over all observers.
	The spectrum stays constant in all magnetic field models and agrees with the shape of the energy spectrum observed in many observatories around the globe \citep[cf. e. g. ][]{spectrum,ThePierreAuger:2013eja}.
	Energy loss is due to particle interactions and hence depends on the time that UHECRs travel through ambient photon fields, with whom they interact.
	The difference in deflection between the several magnetic field models is small enough that the travel time of each event is similar, and the recorded energy spectrum is the same. 
	The number of events with the highest energies ($E > 10^{20} \eV$) varies with the observer but not with the magnetic field model, which does not show any effect on the energy spectrum.
	In conclusion, our tests show no statistical footprint of EGMFs in the observed energy spectrum of UHECRs in $10^{18} - 10^{21} \eV$ energy range.  
	Indeed, the observed spectra are consistent within the scatter even with the model without magnetic fields ({\it b0}). 
\subsection{Angular Power spectrum}
\label{subsec:powerspectrum}
\begin{figure*}
\centering
	\includegraphics[width=\onewidth]{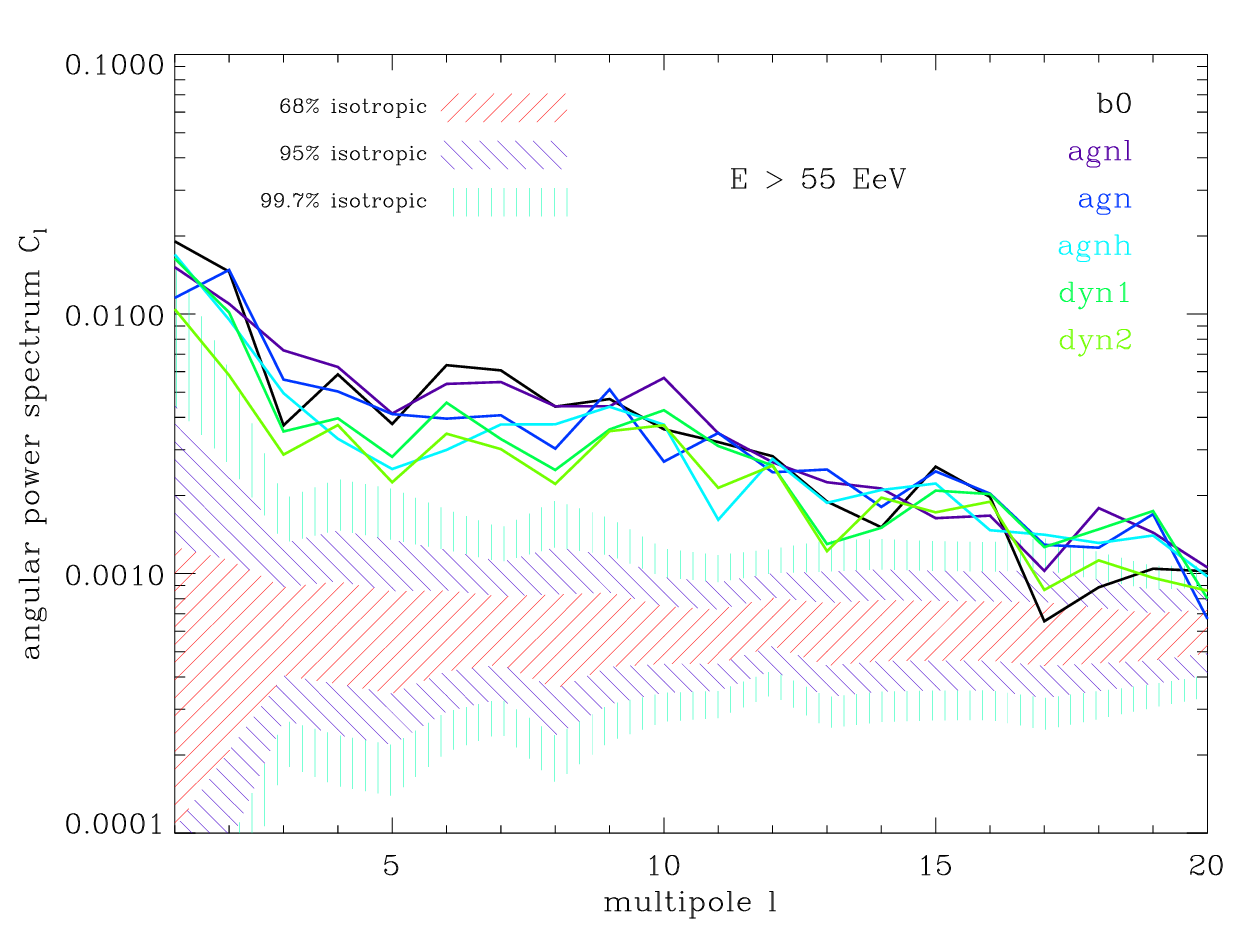}
	\includegraphics[width=\onewidth]{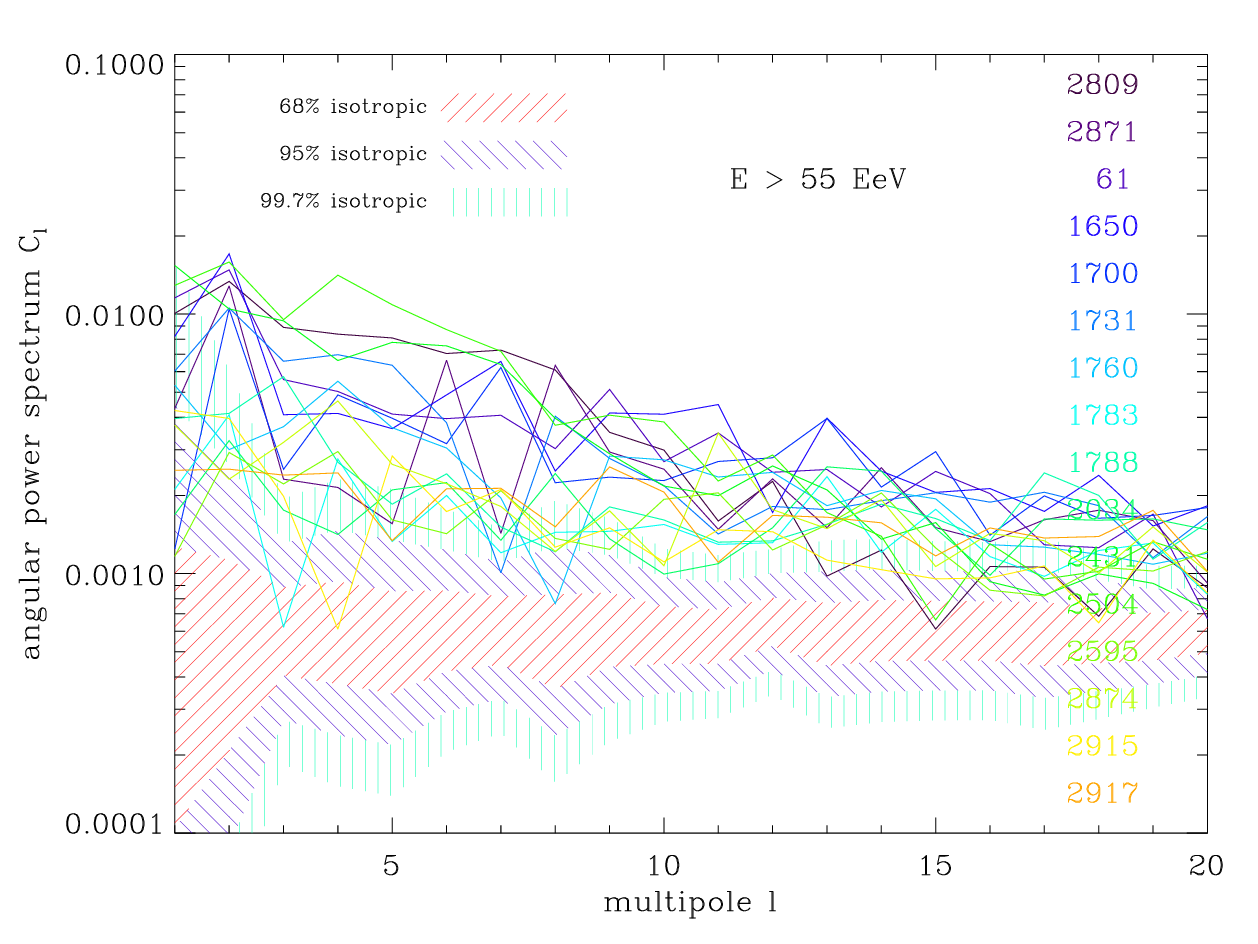}
	\caption{Left: Angular power spectrum of UHECR events observed by ID61 with energies $E > 55$ EeV for the different magnetic field models.
Right: same as left, all 16 observers in one model ({\it \simulation}). }
	\label{fig:powerspectrum61}
	\label{fig:autospectrum1}
\end{figure*}
	To compute the angular power spectrum presented in this section, we first produced a full-sky map of the arrival directions of the UHECR events as seen by the observer (e. g. Fig. \ref{fig:fullsky_events}).
	From this we computed the spherical harmonics and the angular power spectrum, using the {\it anafast} procedure of the {\it healpix}\footnote{http://healpix.jpl.nasa.gov/} package \citep[e. g.][]{arXiv:1411.2486v2}.
	Finally, the whole angular power spectrum was normalized by the total flux.
	We produced 30 runs for the isotropic prediction, where we let CRPropa inject the $10^7$ particles on random positions throughout the simulation volume, resulting in a comparable number of observed events.
	The shaded areas show the 1, 2 and 3$\sigma$ standard deviation of these isotropic runs.
\\ \\
	We see for the angular power spectrum in Fig. \ref{fig:powerspectrum61} that the predicted anisotropy is significant to more than $99 \%$ for all multipoles $\leq 20$.
	This disagrees with results from experimental observations by \citet{powerspectrum}, which suggest that only the dipole moment shows a significant excess from isotropy while
	higher-order moments are in line with the isotropic prediction.
	In the real case, the significantly larger anisotropy detected on average by our set of observers might be erased by the additional deflection from the intervening magnetic fields of the Milky Way. 
	These results are in line with  \citet{arXiv:1411.2486v2}, who used the real observed distribution of galaxies in the 2MASS survey and predicted the power spectrum of observed UHECRs events assuming a typical smearing around sources due to EGMFs.
	They also reported that large-scale structures should produce a significant excess of high order moments in the power spectrum of UHECRs, compared to the isotropic (and observed) prediction.
	They argued that, while the obliteration of high-order moments by the magnetic field of the Milky Way might be ruled out, additional isotropisation can be due to large-scale magnetised winds around galaxies, acceleration of CRs outside of galaxies, or to an increased acceleration activity by the distant (and hence more isotropic) distribution of galaxies.
	However, the regular component of the galactic magnetic field is yet not known well enough to make very accurate predictions \citep[e. g.][and references therein]{janssonfarrar1,janssonfarrar,Beck:2014pma}.
\\ \\
	The comparison of Fig. \ref{fig:powerspectrum61} with Fig. \ref{fig:source_histo} shows that a strong dipole moment ($\lesssim 10^{-2}$) coincides with a dominant source within 5 Mpc distance to the observer (e. g. 61, 1650, 2431, 2504, 2809).
	With this in mind, Fig. 4 in \citet{powerspectrum} hints to a dominant nearby source, probably Centaurus A.
\\ \\
	We also notice that a minority of our observers in Fig. \ref{fig:autospectrum1} (right panel, e. g. 1783, 2034, 2595, 2915, 2917) actually measure a distribution of  UHECRs which is consistent with isotropy at the $2-3 \sigma$ level for most of the multipoles.
	This stresses that the local ($\ll 50 ~\rm Mpc$) environment is crucial in generating the observed patterns of UHECRs as it can influence the observed (an)isotropy of events more significantly than the tested magnetic field models.
\\ \\
\begin{figure*}
\centering
	\includegraphics[width=\onewidth]{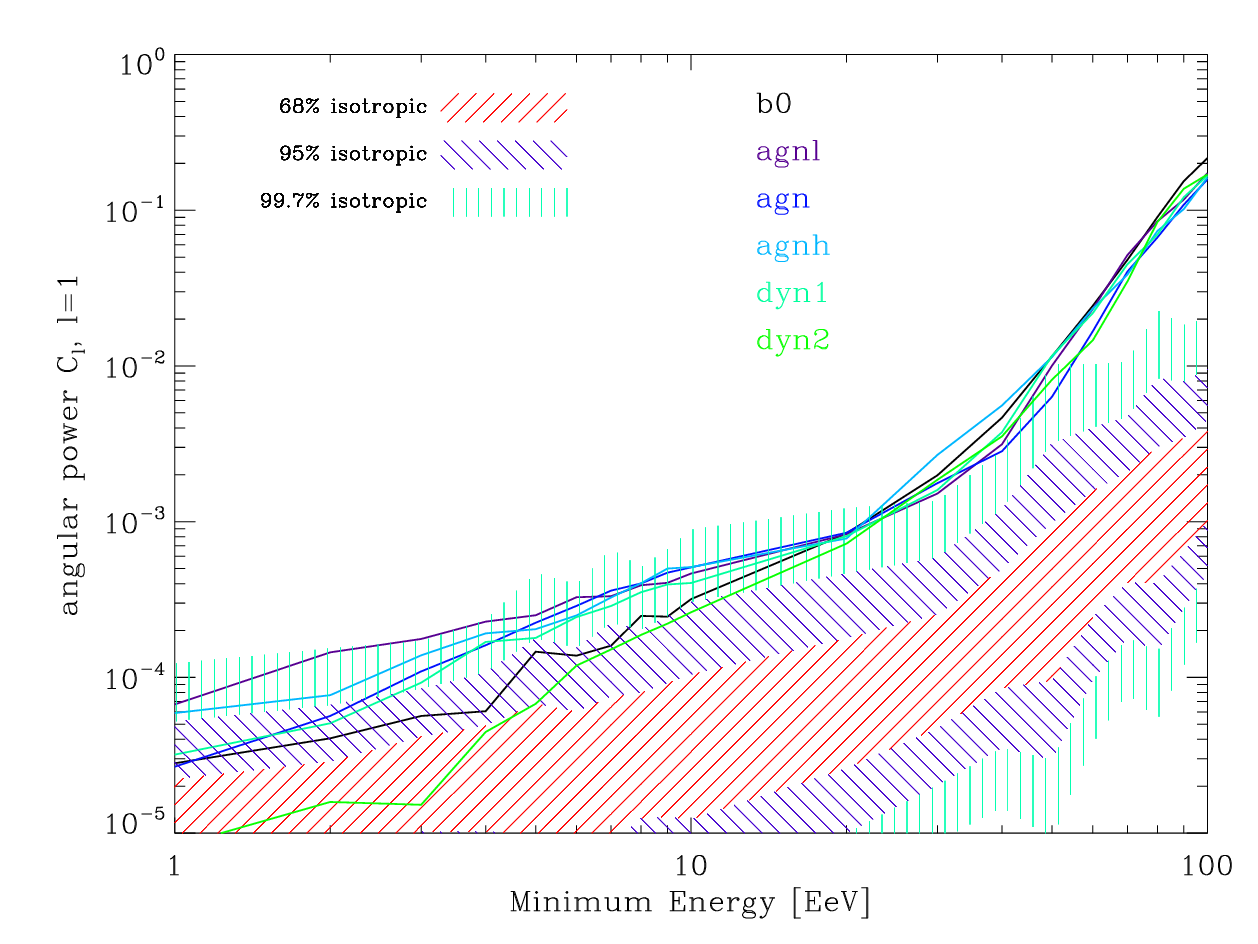}
	\includegraphics[width=\onewidth]{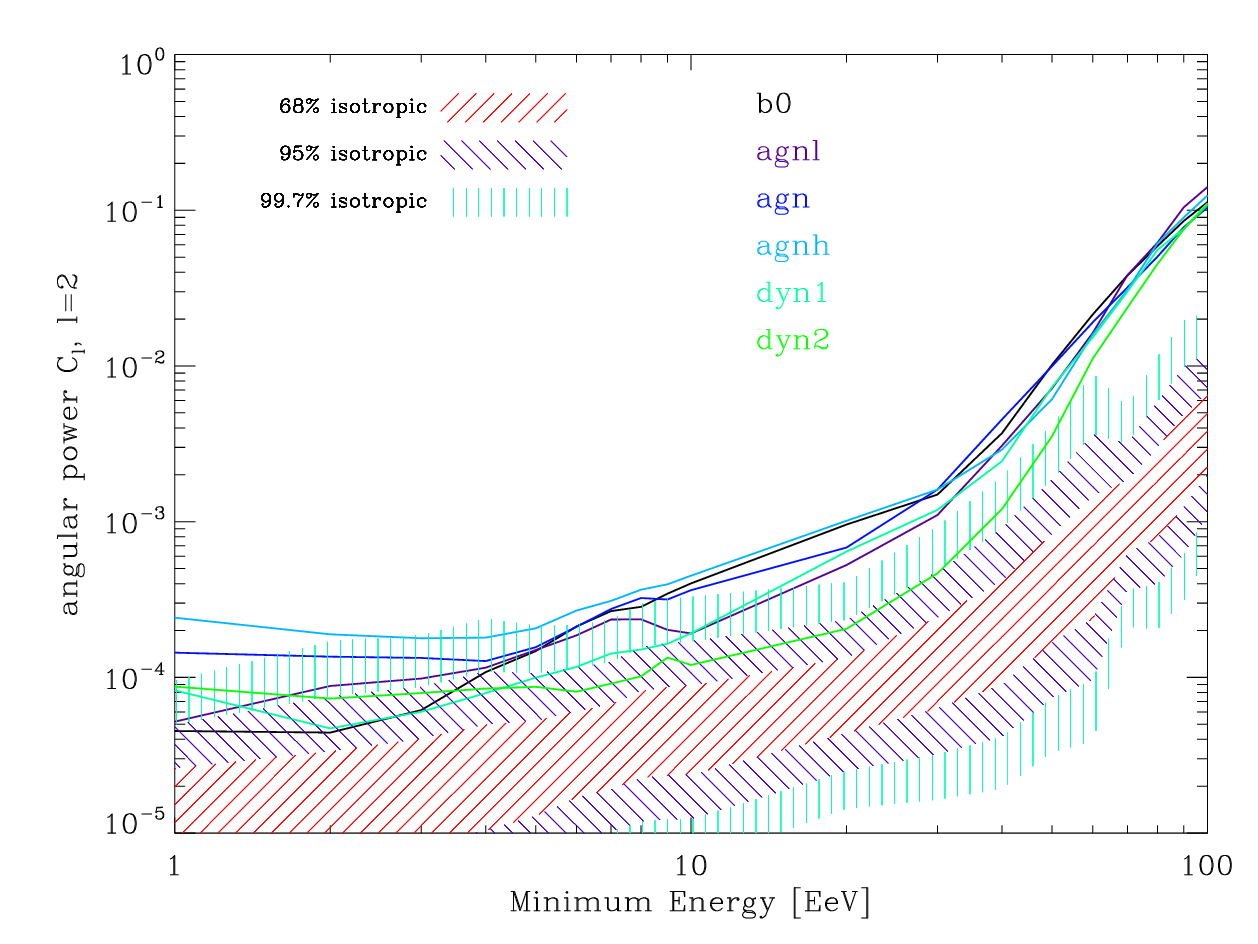}
	\caption{Angular power for the first two multipoles as function of minimum energy of UHECR events observed by ID61.} 
	\label{fig:energy_function}
\end{figure*}
	In Fig. \ref{fig:energy_function} we show the energy dependence of the angular power of the dipole and quadrupole moment,
	both increase with increasing minimum energy.
	The slope of this dependence steepens at $40 \EeV$, which is the GZK-threshold.
	This shows that the excess from isotropy, which occurs at higher energies, is due to the GZK-effect that confines UHECR travel to the local universe.
	On the scale of the GZK-horizon of roughly $100 \Mpc$, matter is not distributed isotropically, and hence an anisotropic signal of extragalactic UHECRs is expected \citep[e. g.][]{GZK_horizon1,LSS}. 
\\ \\
\begin{figure*}
\centering
	\includegraphics[width=\onewidth]{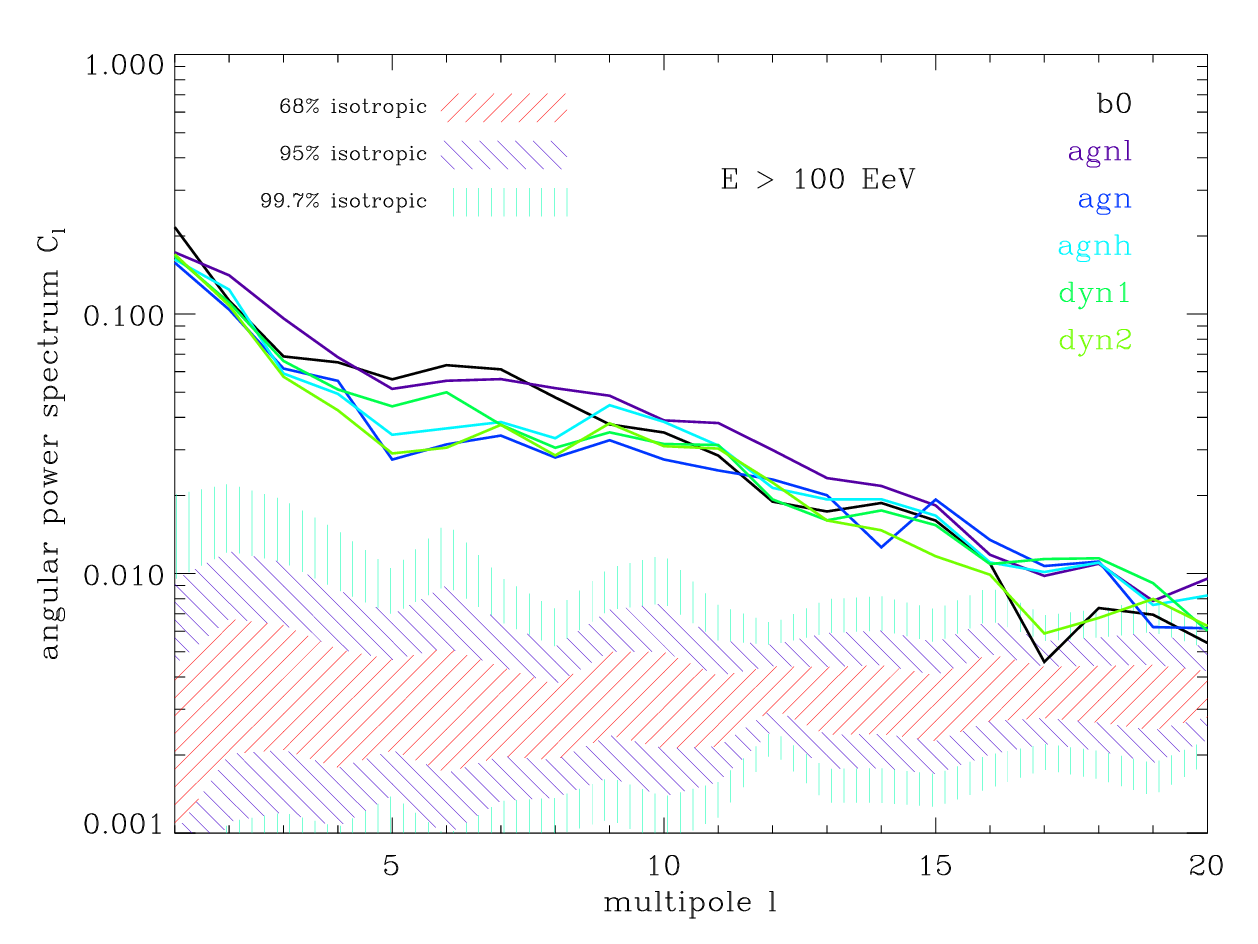}
	\includegraphics[width=\onewidth]{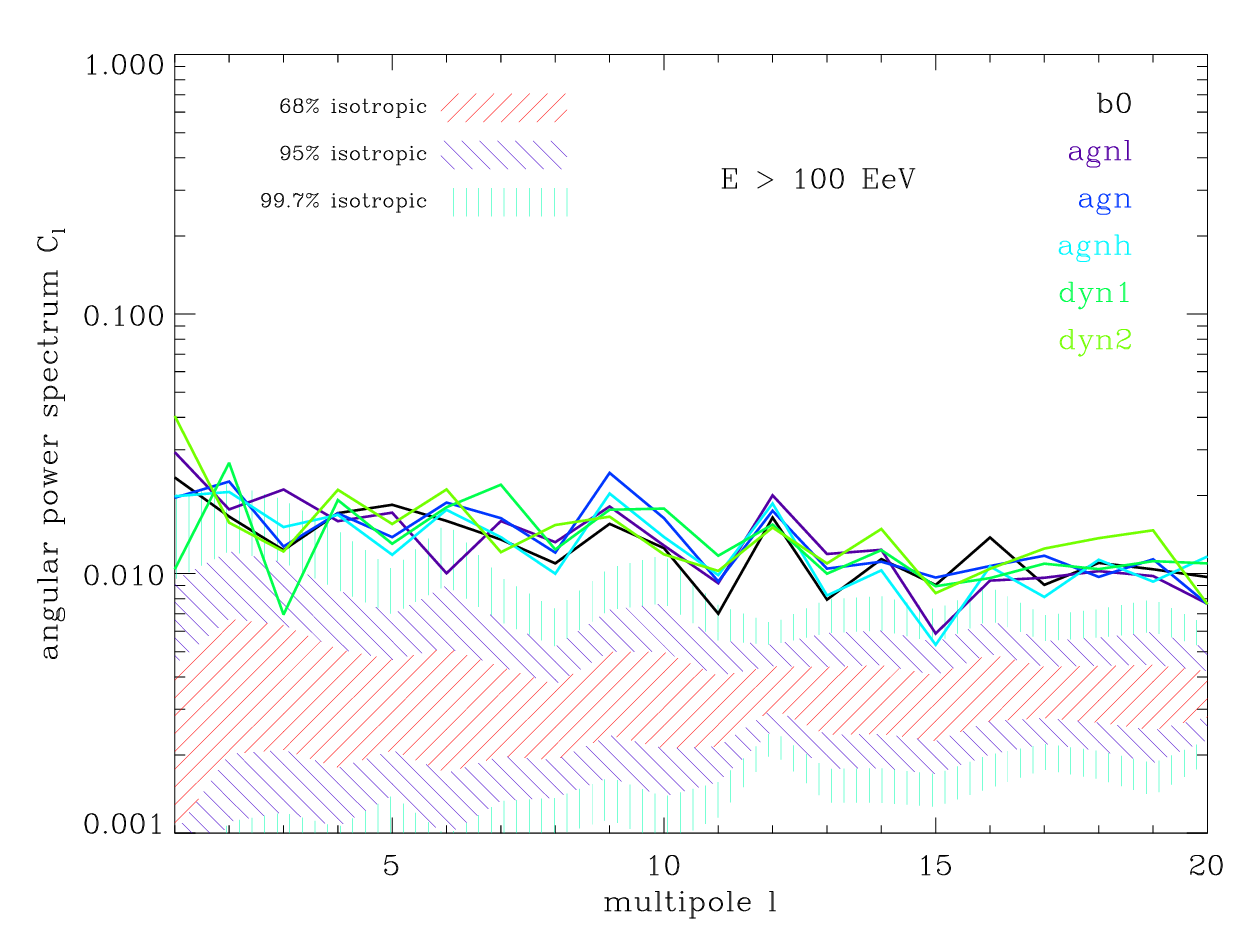}
	\caption{Same as Fig. \ref{fig:powerspectrum61}, left, for minimum energy $E > 100 \EeV$. Here we compare the angular power spectrum for an observer with a source in 4 Mpc distance (ID61, left) and without any source within 10 Mpc distance (ID2917, right).} 
	\label{fig:extreme_cases}
\end{figure*}
\begin{figure}
\centering
	\includegraphics[width=\onewidth]{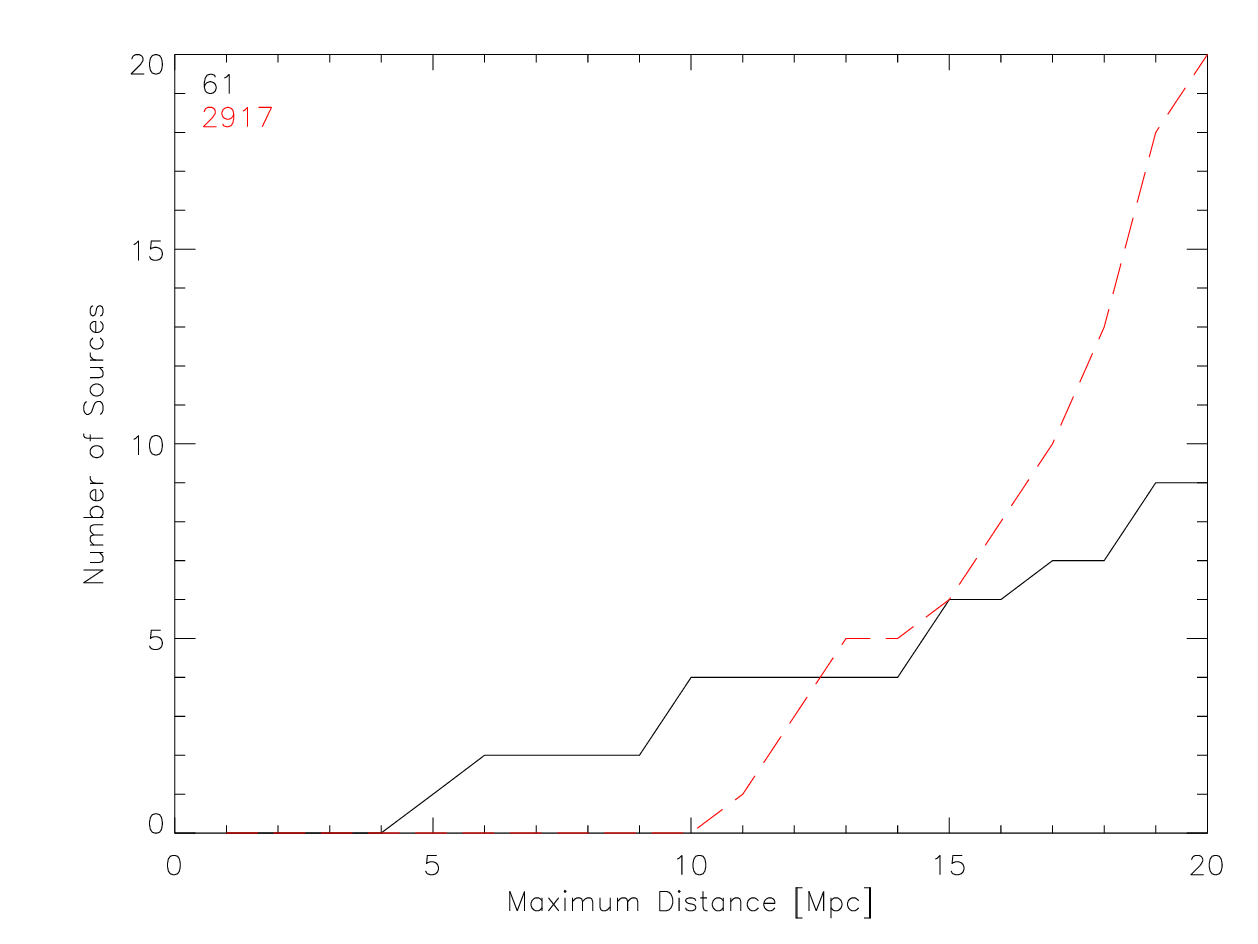}
	\caption{Number of sources within given distance for the observer with a source at 4 Mpc distance (ID61, solid) and the one without any source within 10 Mpc (ID2917, dashed).}
\label{fig:extreme_cases_histo}
\end{figure}
	In Fig. \ref{fig:fullsky2917}, right, we show the full sky of an observer with no sources within $10 \Mpc$ distance (ID2917).
	In Figs. \ref{fig:extreme_cases} and \ref{fig:extreme_cases_histo} we compare this observer to one with a dominant source in $\sim 4 \Mpc$ distance (ID61) that has been presented earlier. 
	In the case of no nearby source, the angular power spectrum shows no significant deviation from isotropy in the lower multipoles.
	If there is a source located close to the observer, the angular power spectrum increases and eventually exceeds the isotropic prediction significantly.
 \\ \\
	In summary, our harmonic analysis suggests that the geometrical location of cosmic sources of UHECRs affects the distribution of arrival directions of events more significantly than  different models of EGMFs.
	Indeed, observers located in the same simulated universes can measure very different power spectra of events, just because of cosmic variance. 
	This effect dominates any dependence of predicted anisotropies on the class of EGMF models studied here, limiting the possibility of inferring the origin of EGMF based on the observed properties of UHECRs.
	It also stresses the importance of a realistic characterisation of the local environment of the observer to interpret observations on Earth. 
	This can be studied further making use of "Constrained Simulations" of the Local Universe \citep[e. g.][]{2013MNRAS.432..894J,2016MNRAS.455.2078S}.

\subsection{Separation Angle}
\label{subsec:separationangle}
	\citet{Auger2010} investigated the anisotropy of the UHECR arrival signal by considering the angular distance of events from Cen A.
	Due to its similarity, all the graphics presented in this section, if not stated otherwise, are shown for observer ID61 whose closest source is at $\sim 4 \Mpc$ distance.
\\
\begin{figure*}
\centering
	\includegraphics[width=\onewidth]{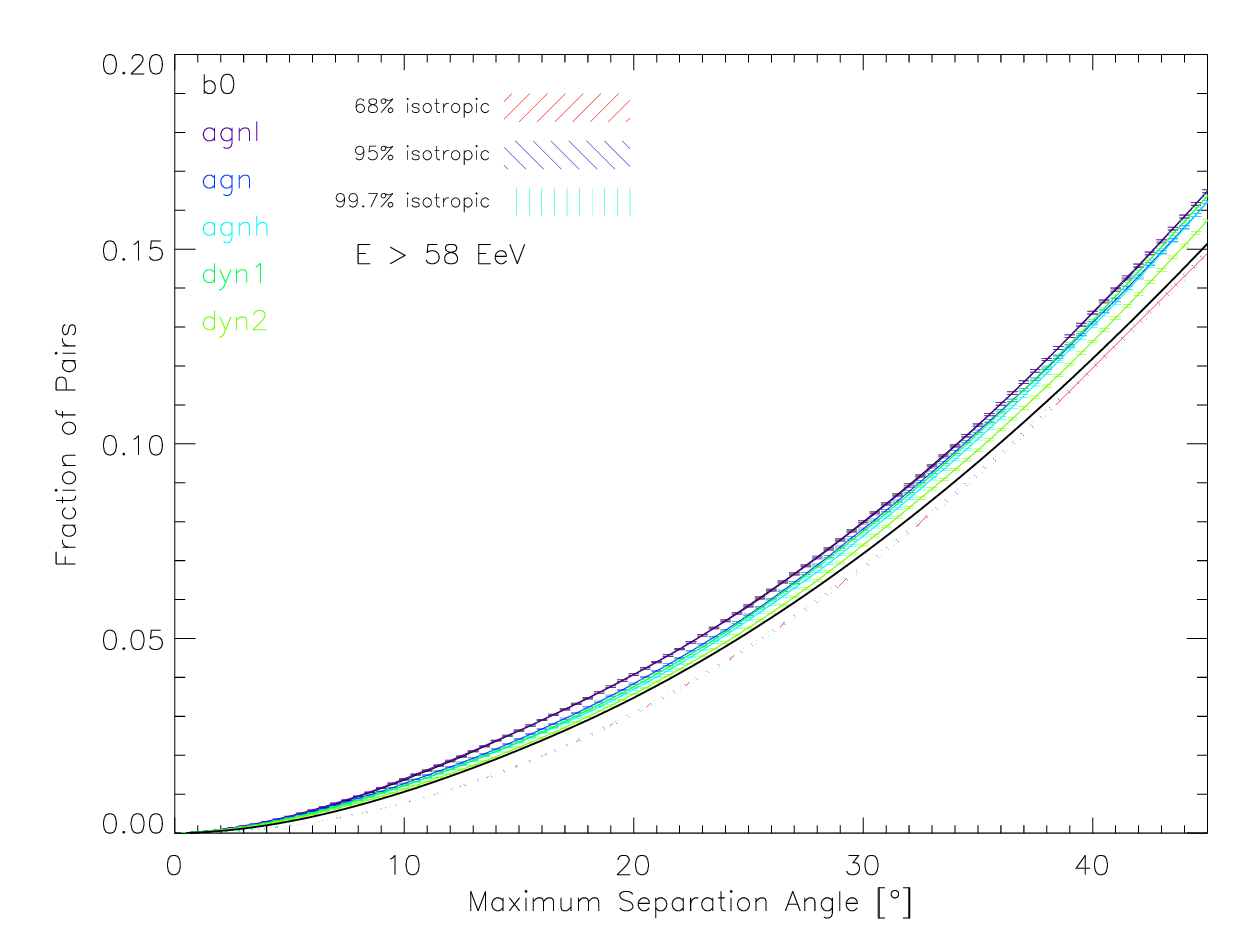}
	\includegraphics[width=\onewidth]{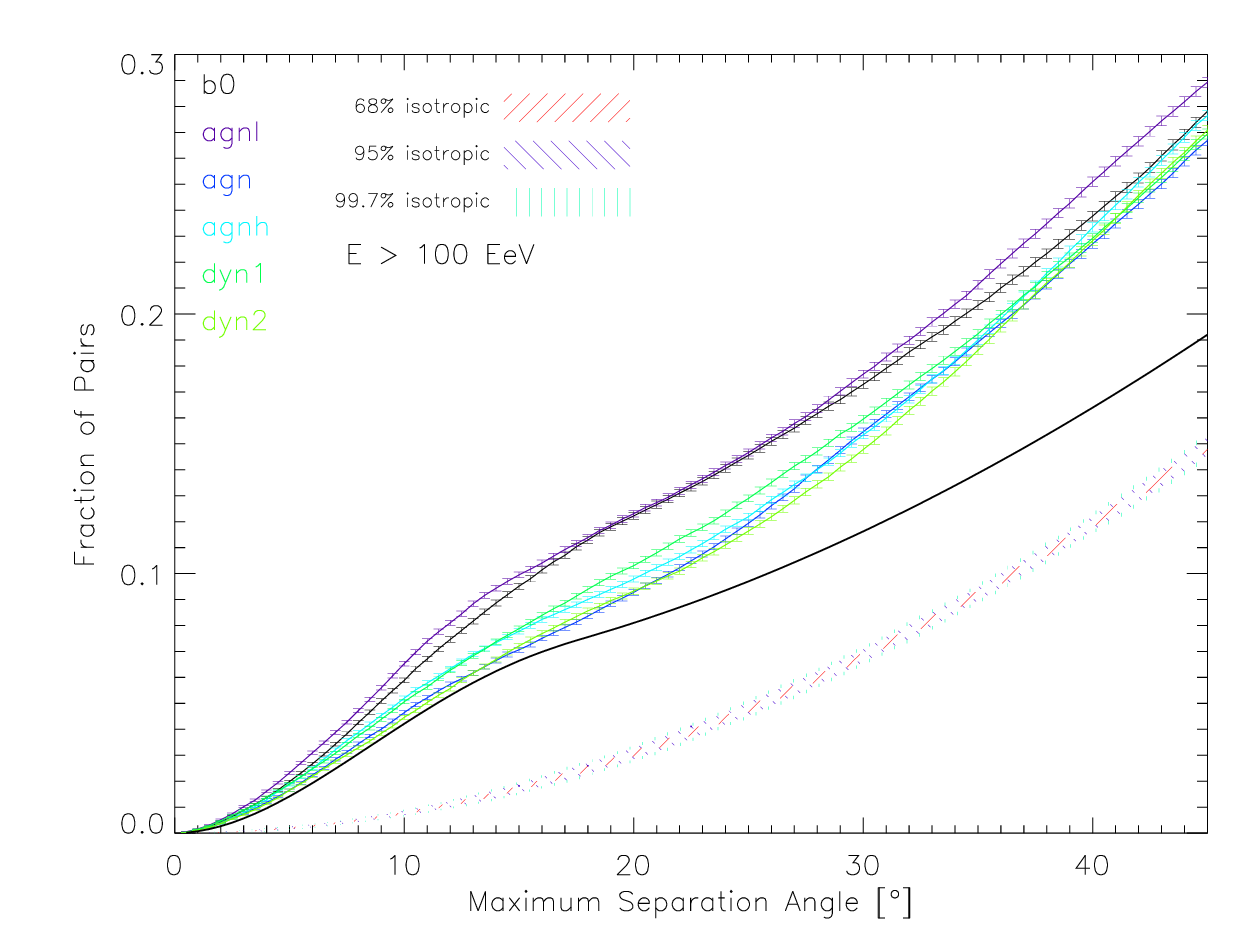}
	\caption{The fraction of pairs of observed events, $E > 58,\ 100 \EeV$ (left, right), separated by less than the given angle. The coloured lines indicate the magnetic field model, the error bars show the Poissonian noise. The thick black line shows the isotropic plus the geometric prediction.}
	\label{fig:pairseparation58}
	\label{fig:pairseparation100}
\end{figure*}
	The separation angle of pairs of events in Fig. \ref{fig:pairseparation58} was computed via the scalar product of their arrival direction, i. e. the normalized negative momentum vector of the UHECRs at time of observation.
	The theoretical prediction for isotropy for the distribution of events over a sphere is given by the cumulative sine function and is in good agreement with the simulated isotropic cases.
	We can also predict the influence of the observer effect (cf. Sec. \ref{subsec:geometriceffect}) for the signal of the closest source.
	To do so, we produced a test simulation without magnetic fields with a single source of UHECRs at a distance comparable to that of the closest source for observer ID61.
	The separation angle function of this test gives our prediction for the finite observer effect for pairs of events that both originated from the closest source.
\\ \\
	The left panel of Fig. \ref{fig:pairseparation58} can be compared with Fig. 8 in \citet{Auger2010}.
	For events above $E > 58\EeV$ there is no significant deviation from isotropy, but for $E > 100 \EeV$ there is a clear excess.
	This is expected because, due to lower average travel distances at higher energies, there is a higher number of events that originated from the closest source.
	The influence of the geometric effect changes the form of this plot by flattening the slope in the low separation part.
	We emphasize that this actually affects all events from close-by sources, while our geometric prediction only considers those from the closest source.
	Nonetheless, there is a significant difference between the magnetic field models:
	the magnetic field free case ({\it b0}, black) and the simulation with the lowest magnetisation ({\it agnl}, purple, cf. Fig. \ref{fig:Bhisto}) show the strongest deviation from isotropy.
	However, for other observers the case is different and the presence of nearby sources can cause strong anisotropies even
	in models with higher magnetic fields. 
	On the other hand, if no sources are present close to the observer,  the differences between the magnetic field models are very small.
\\ \\
\begin{figure*}
\centering
	\includegraphics[width=\onewidth]{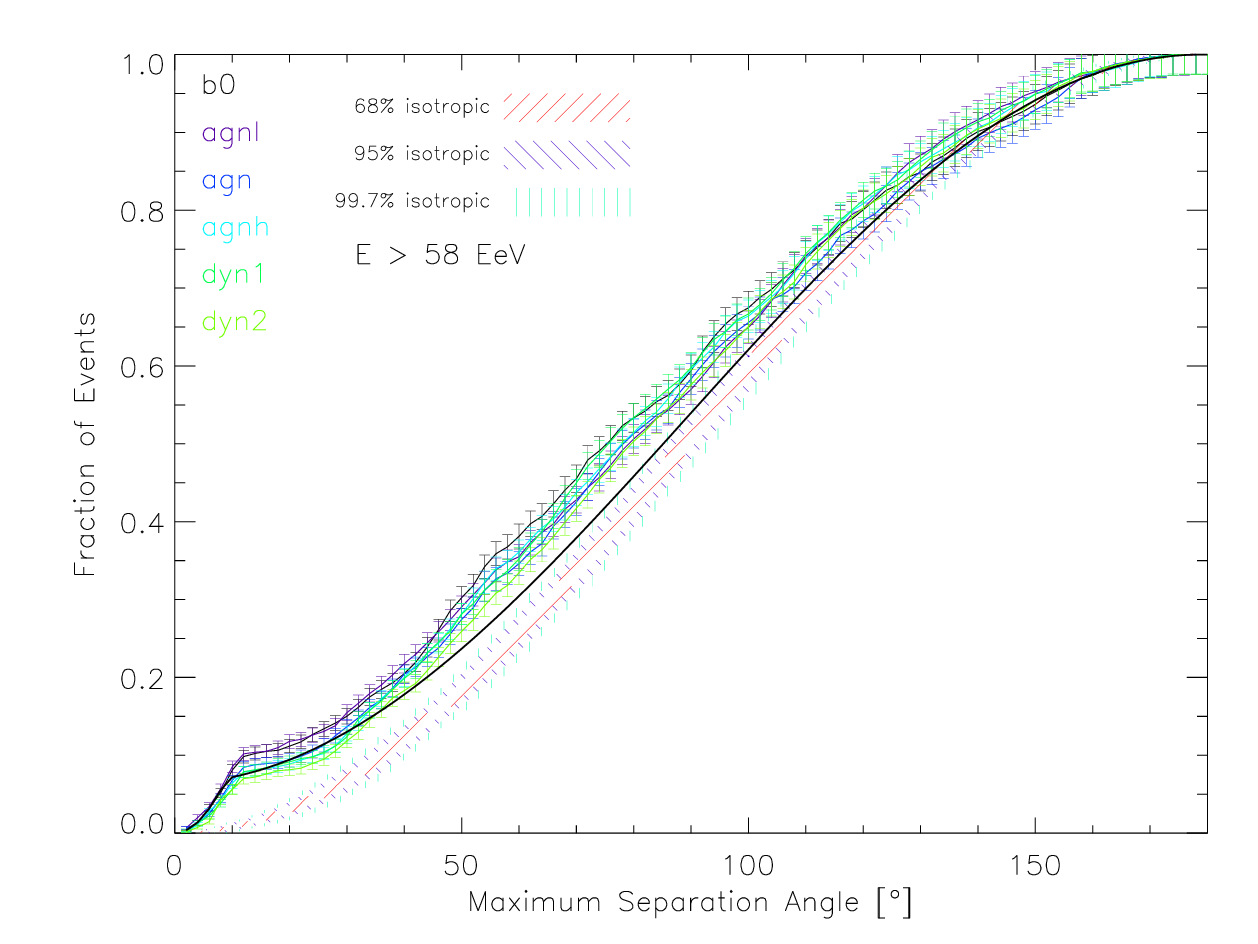}
	\includegraphics[width=\onewidth]{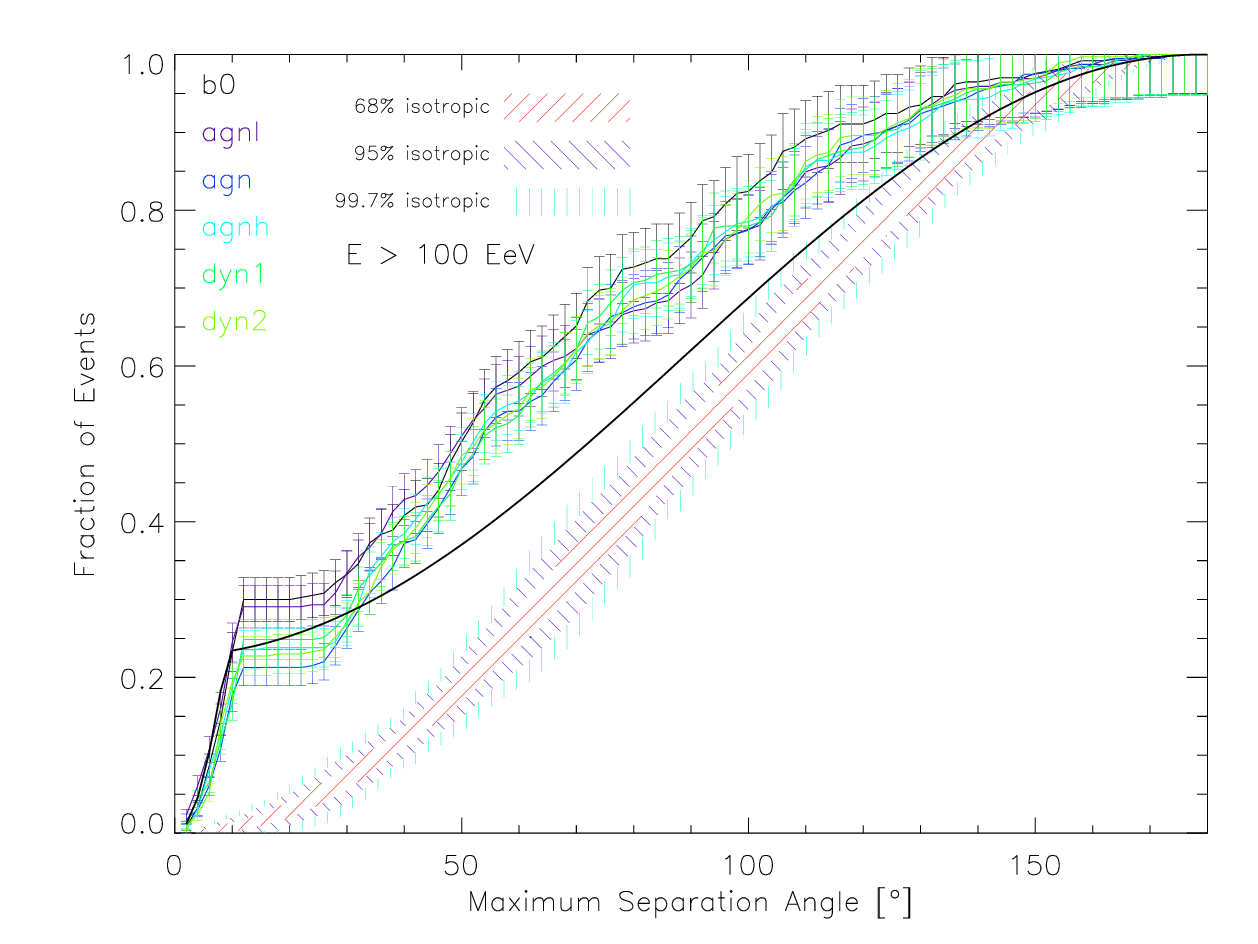}
	\caption{The fraction of observed events, $E > 58, 100 \EeV$ (left, right), with a separation angle to the closest source less than the given angle. The coloured lines show the signal of the different magnetic field models, the error bars show the Poissonian noise. The thick black line shows the isotropic plus the geometric prediction.}
	\label{fig:separationangle58}
	\label{fig:separationangle100}
\end{figure*}
\begin{figure}
\centering
	\includegraphics[width=\onewidth]{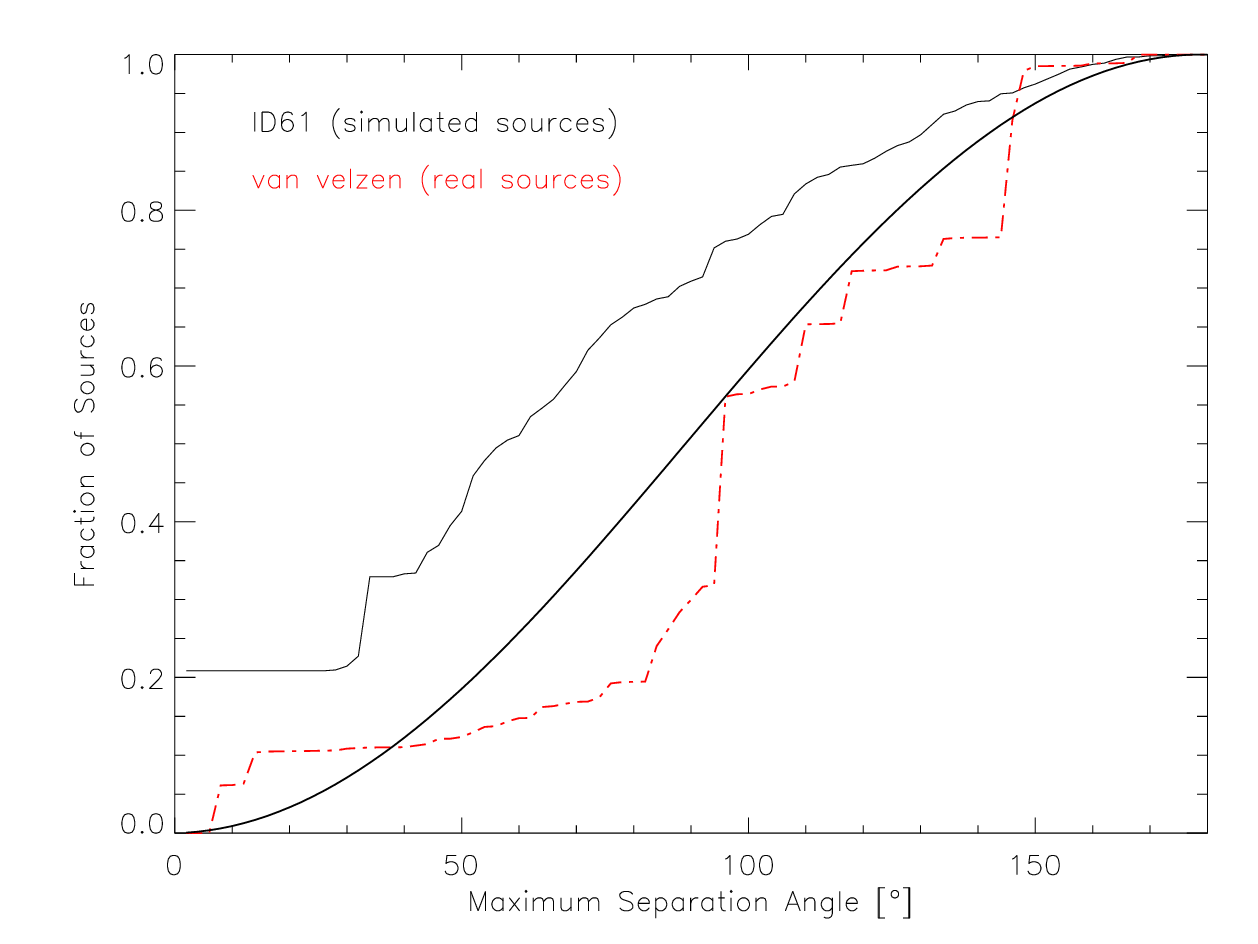}
,	\caption{Same as Fig. \ref{fig:separationangle58}, but here we calculate the separation angle between the closest source and all other sources.
The result is divided by the distance squared.
Shown is the separation angle of sources for simulated observer ID61 (solid) and for the van Velzen catalogue (dash-dot) of real radio sources  with respect to the position of Cen A.}
	\label{fig:separationanglesources}
	\label{fig:histovelzen}
\end{figure}
	The separation angle from the closest source in Fig. \ref{fig:separationangle58}  was computed via the scalar product of the arrival direction of events with the vector that points from the centre of the observer to the source with smallest distance to the observer.
	The left panel of this Figure can be compared to Fig. 10 of \citet{arxiv:1411.6111v3} which shows the distribution of separation angles from Cen A.
	The isotropic prediction is the same as in Fig. \ref{fig:pairseparation58}, a cumulative sine function plus the prediction for the geometric effect, which is here normalised to the number of observed events from the closest source.
	 In all magnetic field models, the signal at small angles is found to be dominated by the finite geometrical size of the observer.
\\ \\
	In addition to the contribution of the closest source to the small angle part of the separation angle function, in Fig. \ref{fig:separationangle58}  there is also a slight excess from isotropy at $\sim 50^\circ$, which becomes stronger for $E > 100 \EeV$.
	In the full-sky maps in Fig. \ref{fig:fullsky_events} an excess in the number of events can be seen around other very close-by sources.
	This suggests that the sharp increases in Fig. \ref{fig:separationangle58} are due to the separation angles between the closest source and another very close-by source.
	Indeed, Fig. \ref{fig:separationanglesources} shows that the anisotropy signal sharply increases where there is a crowd of very nearby sources with matching angular distance to the closest source (even if the finite observer effect of Sec. \ref{subsec:geometriceffect} can significantly smear out this feature). 
	We suggest that this property can be used
 to correlate catalogues of possible extragalactic sources with UHECR observations in order to identify them as sources of UHECRs.
\\ \\
	Figs. \ref{fig:pairseparation_simulation} and \ref{fig:separationangle_simulation} show the separation angle plots for all observers in the {\it \simulation} \ model.
	Again cosmic variance has a higher impact than the magnetic field model.
\\
	The separation angles of pairs, Fig. \ref{fig:pairseparation_simulation}, are dominated by the most nearby sources.
	For the plot of separation angles between events and the closest source in Fig. \ref{fig:separationangle_simulation},  the small angle part is dominated by the closest source, with an excess compared to the isotropic distribution below $10^\circ$. 
	Larger angles are instead dominated by the other nearby sources.
\\
	In summary, for an observer with a dominant source at approximately the same distance as Cen A we find that the angular separation of $E>58 \EeV$ events is consistent with isotropy for all magnetic field models, and that the level of anisotropy increases with increasing minimum energy.
	On the other hand, for $E>100 \EeV$ events we find significant departures from isotropy, which can be traced back to the clustering of sources seen in projection. \\ \\
	
\begin{figure}
\centering
\includegraphics[width=\onewidth]{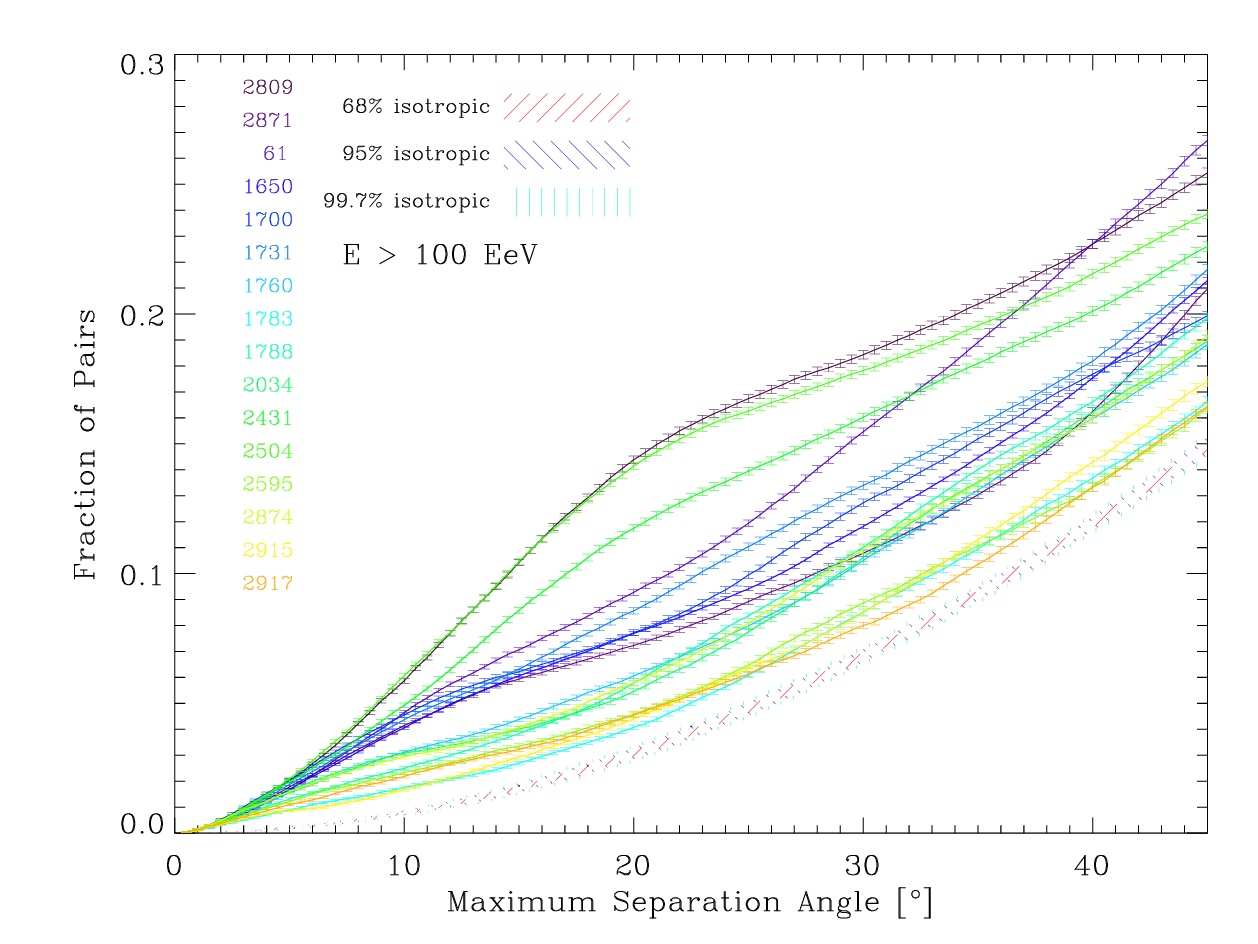}
\caption{Same as Fig. \ref{fig:pairseparation58}. Shown is the separation angle plots for all observers in the \simulation \ model.}
\label{fig:pairseparation_simulation}
\end{figure}
\begin{figure}
\includegraphics[width=\onewidth]{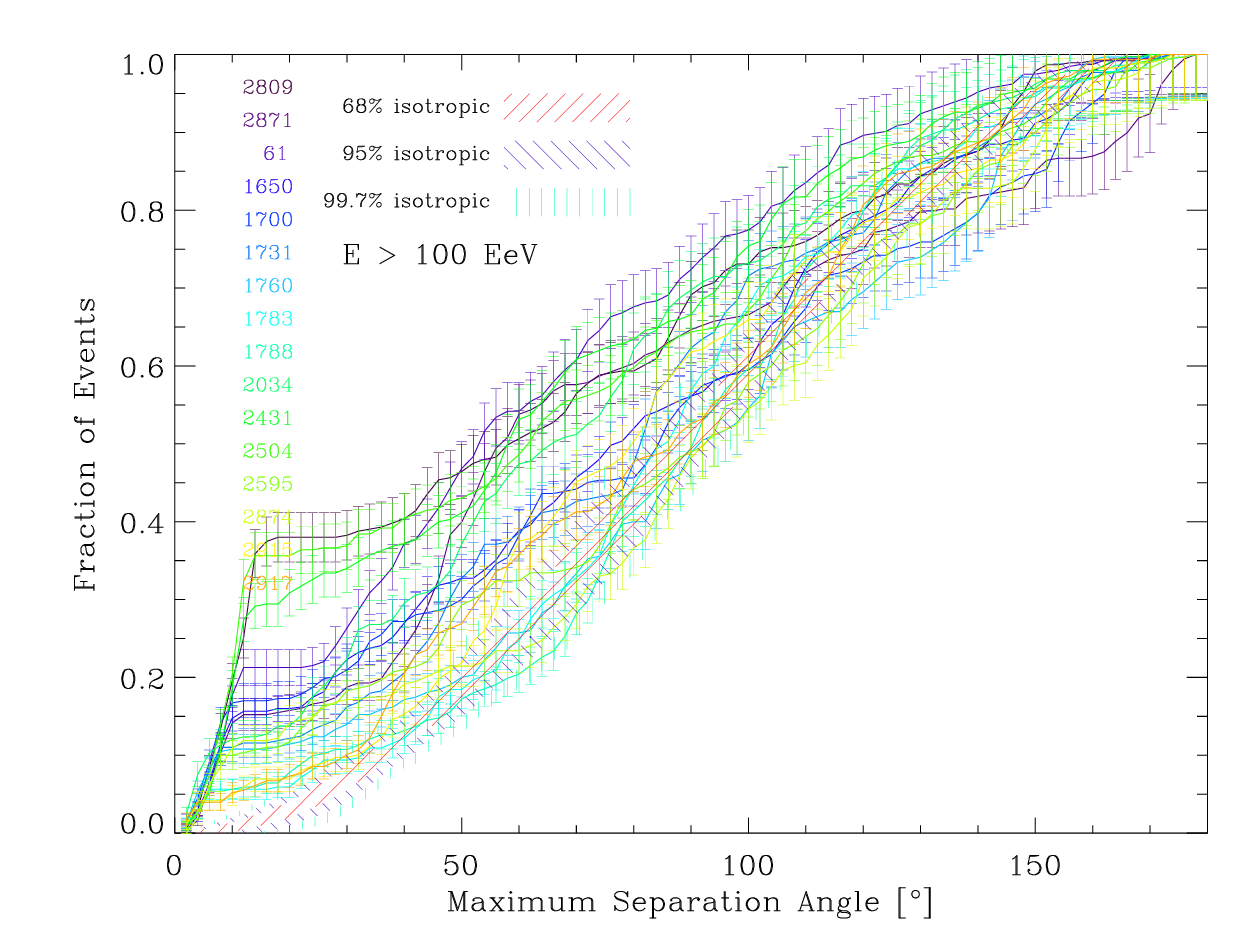}
\caption{Same as Fig. \ref{fig:separationangle58}. Shown is the separation angle plots for all observers in the \simulation \ model.}
\label{fig:separationangle_simulation}
\end{figure}

\subsection{Limiting the strength of magnetic fields in voids}
\label{subsec:plusB}
\begin{figure}
\centering
	\includegraphics[width=\onewidth]{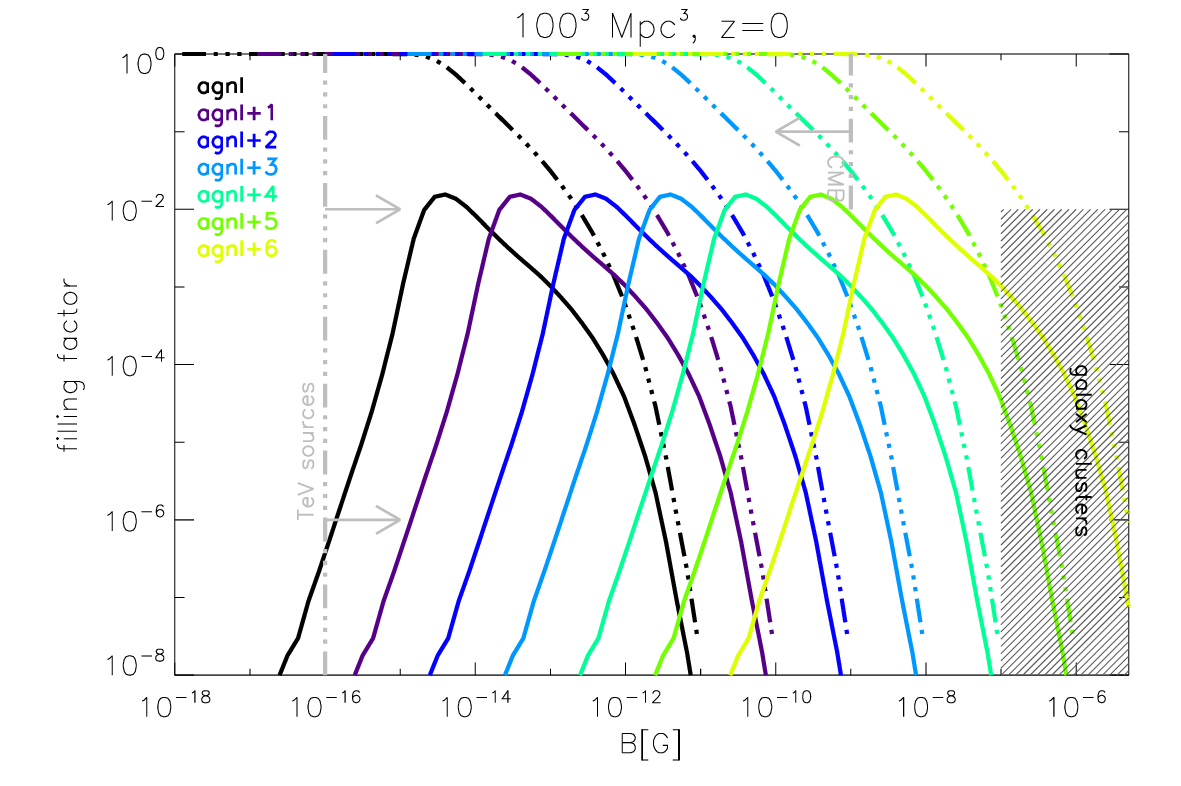}
	\caption{Analogous to Fig. \ref{fig:Bhisto}. Shown is the cumulative and differential (rescaled by 0.1 for clarity) volume filling factor at $z=0$ for the rescaled magnetic fields.}
	\label{fig:plusBhisto}
\end{figure}
\begin{figure*}
\centering
	\includegraphics[width=\onewidth]{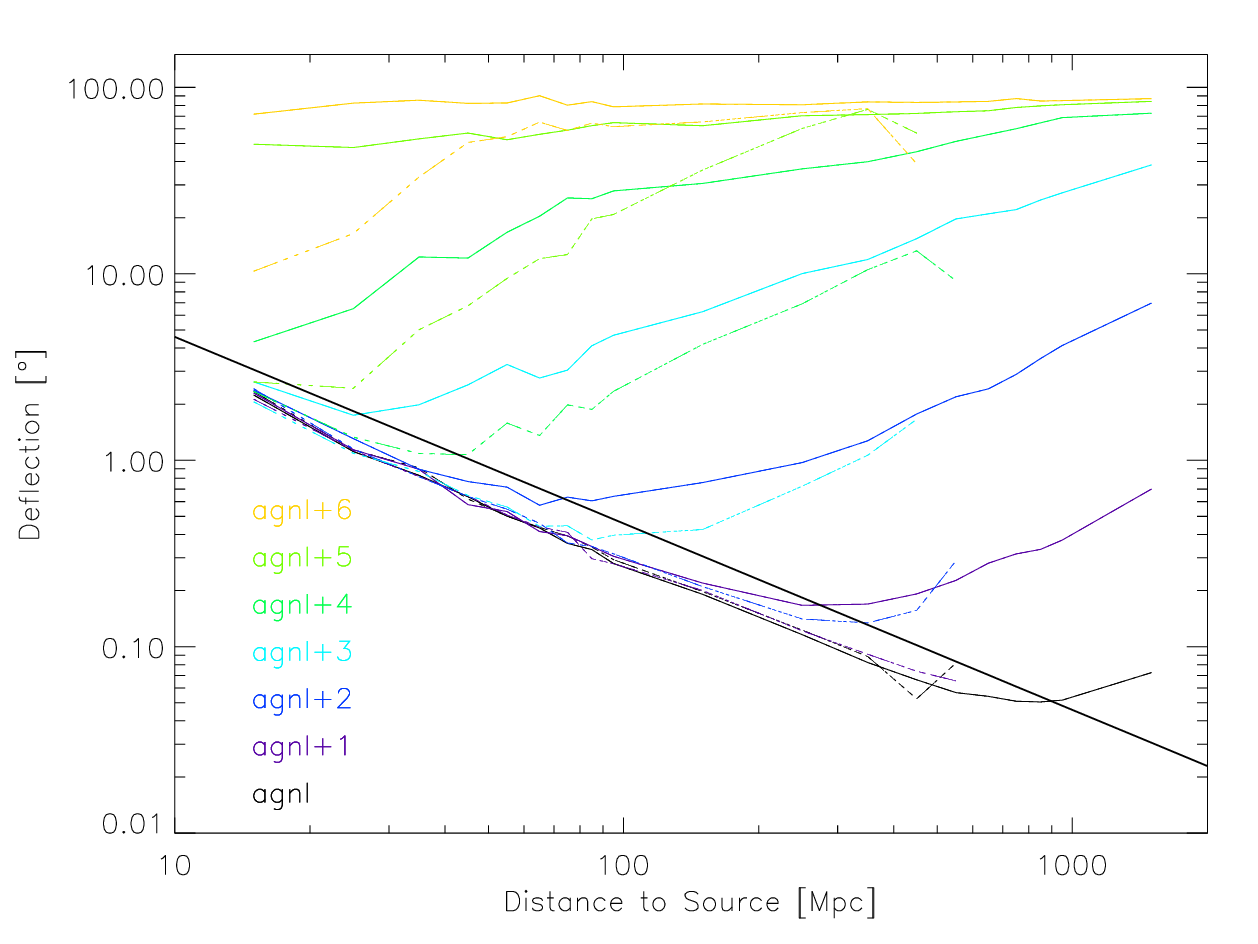}
	\includegraphics[width=\onewidth]{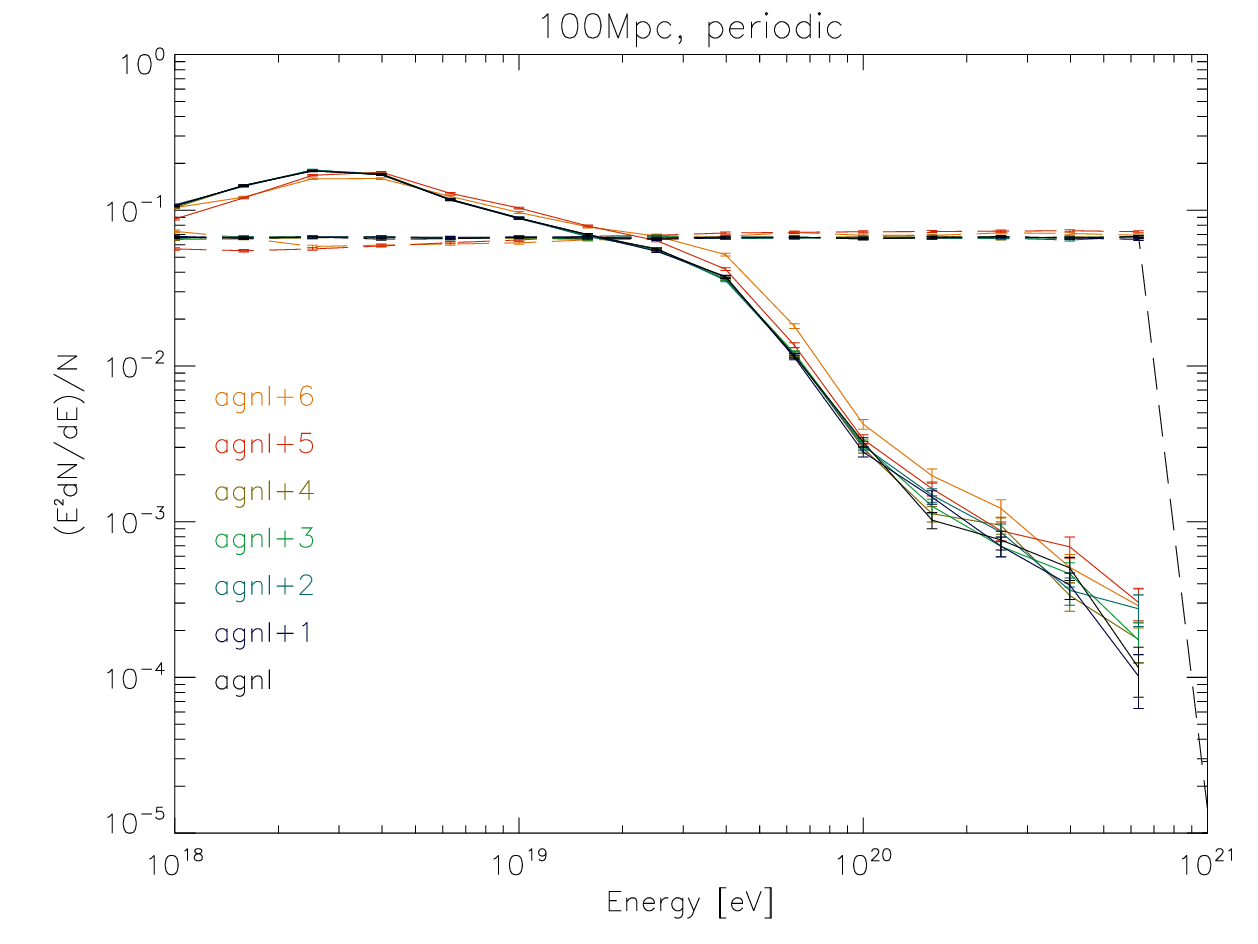}
	\caption{Left: Same as Fig. \ref{fig:deflection_angle} for the rescaled magnetic field models. The average deflection angle of UHECRs above 1 EeV (solid) and 58 EeV (dashed) in the rescaled magnetic field models. 
The black line is the prediction of the maximum artificial deflection angle (cf. Sec. \ref{subsec:geometriceffect}).
Right: Same as Fig. \ref{fig:energyspectrum}. Energy spectrum of observed UHECR events for one observer in the rescaled magnetic field models. The dashed line shows the injected, the solid line the received spectrum. The error bars show the Poissonian noise.}
	\label{fig:deflection_plusB}
	\label{fig:energy_spectrum_plusB}
\end{figure*}
	Here we discuss the influence of the magnetisation of voids on the observable properties of UHECRs.
	To this end, we globally rescaled the magnetic field strength of the {\it agnl} model in steps of factors 10, from $10^1$ to $10^6$, as explained in Sec.\ref{subsec:ENZO}.
	Fig. \ref{fig:plusBhisto} shows the cumulative and the differential volume filling factor of EGMFs for each model, which we used
	to generate UHECR events with CRPropa, limited to the same ID61 observer considered before. 
\\ \\
	The average deflection angle of UHECRs as a function of the distance to their sources is shown in the left panel of Fig. \ref{fig:deflection_plusB}.
	At $E > 1 \EeV$ the deflection of UHECRs from within 100 Mpc is negligible ($< 1^\circ$), until the average field strength in voids is $\sim 10^{-3} \rm nG$ ({\it agnl+2}). For energies $E > 58 \EeV$ the deflection of those UHECRs is negligible even up to  $10^{-2} \rm nG$ ({\it agnl+3}).
	For stronger fields we do find a significant deflection, consistent with earlier results \citep[e. g.][]{Sigl2004,2006ApJ...639..803T}.
\\
	In the right panel of Fig. \ref{fig:energy_spectrum_plusB}, we show the energy spectra measured in this last set of models, which do not show significant differences. 
	However, for $E > 58 \EeV$ we find a slightly increased isotropy when stronger fields are assumed, in both kinds of separation angle plots presented in Sec. \ref{subsec:separationangle}.
	At this energy, the difference in deflection between models is limited to events from sources within 20 Mpc, while for larger energies the difference is further reduced. 
\\ \\
\begin{figure*}
\centering
	\includegraphics[width=\onewidth]{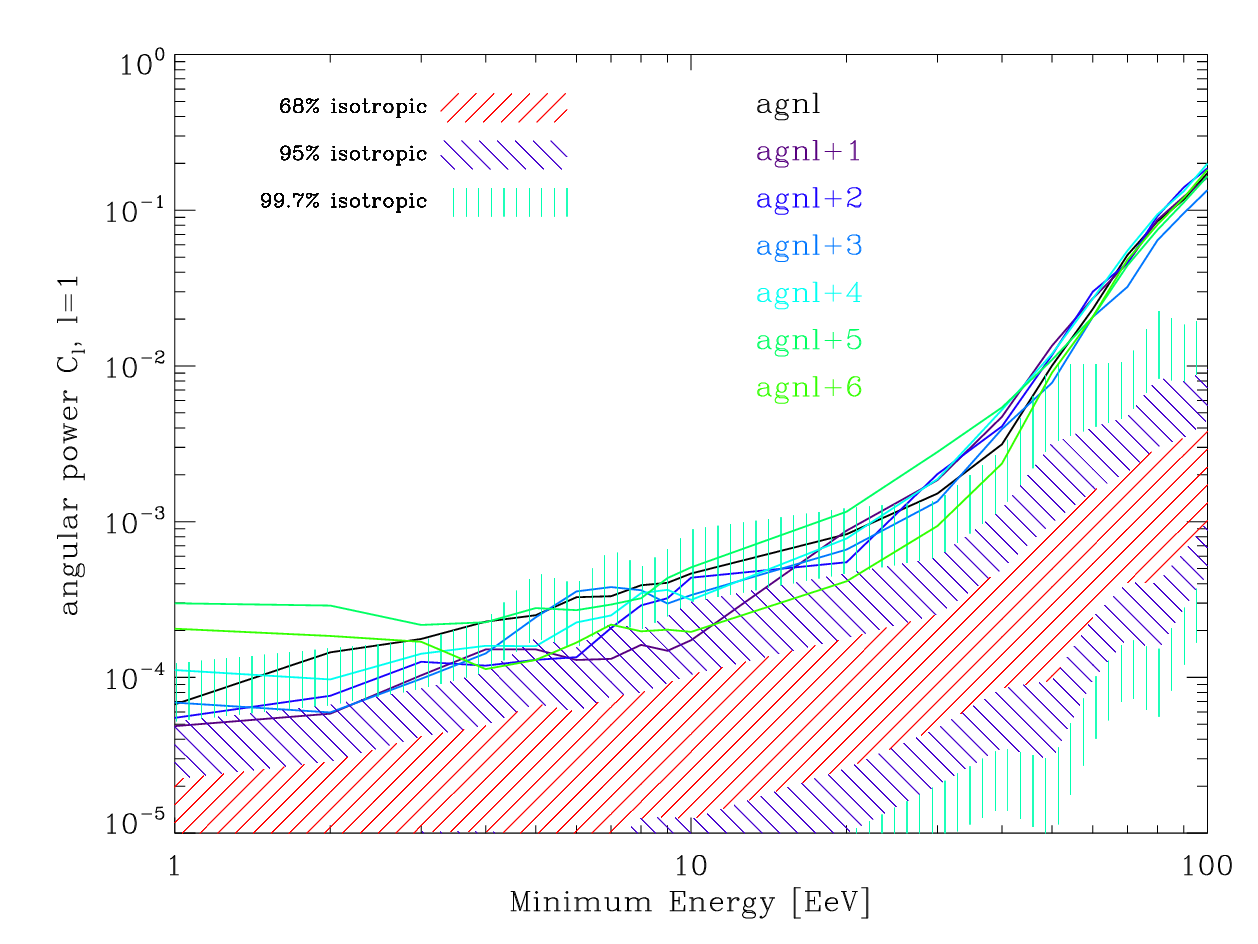}
	\includegraphics[width=\onewidth]{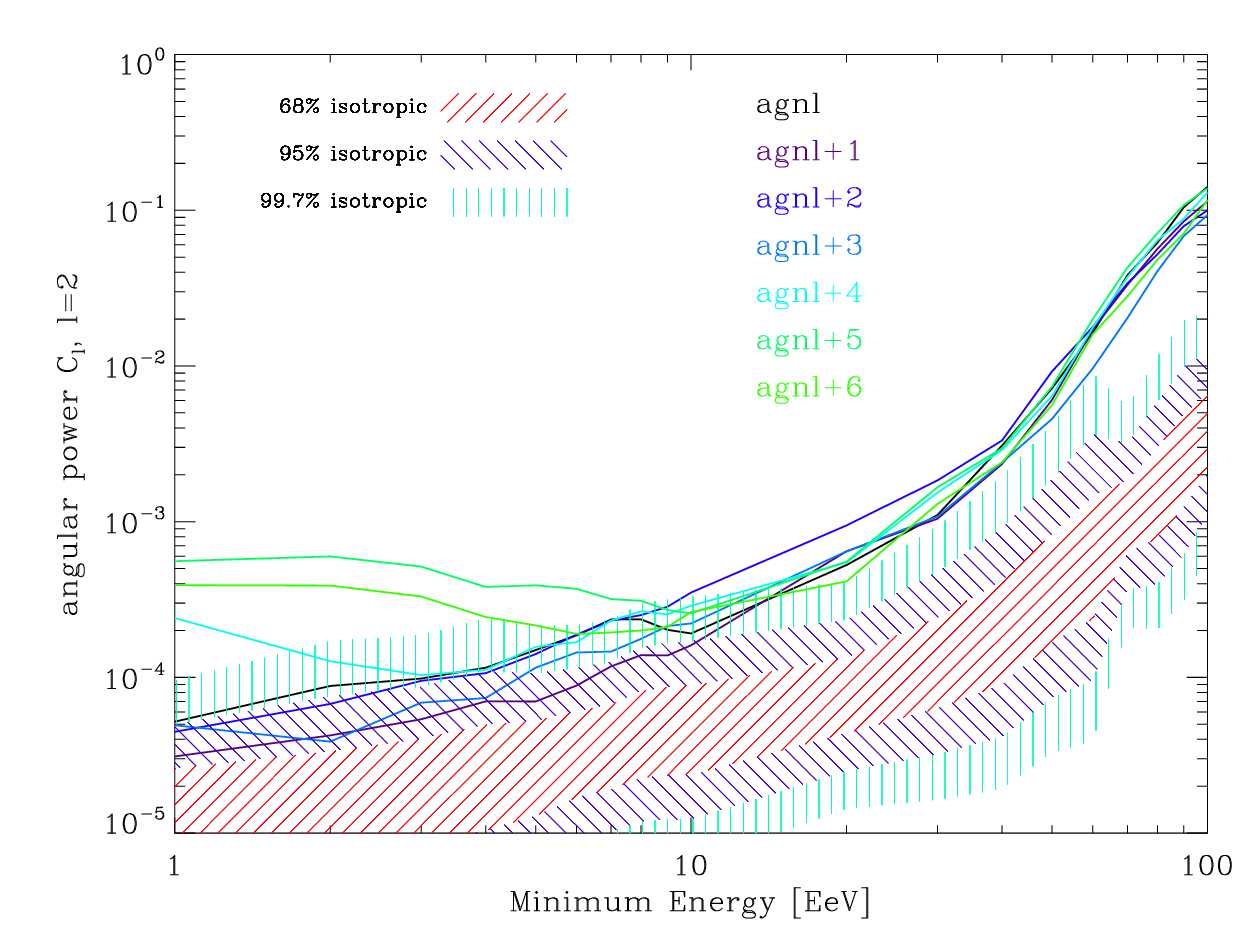}
	\caption{Same as Fig. \ref{fig:energy_function} for the rescaled magnetic field models. Angular power of observed UHECR events for the first two mulitpoles as function of minimum energy of considered particles.}
	\label{fig:efunction_plusB}
\hspace{1cm}
\end{figure*}
	The energy dependence of the two lowest multipoles of angular power in Fig. \ref{fig:efunction_plusB} shows no difference between the different EGMFs  for energies above the GZK-threshold, $E > 40 \EeV$. However, we detect an excess from isotropy at energies of a few EeV, for  $l=1$ in primordial magnetic seed fields of $\geq 1 \ \rm nG$   and for $l=2$ seed fields with $\geq 0.1 \ \rm nG$.
	However, at these energies the observed signal of UHECRs is consistent with isotropy \citep[e. g.][]{ThePierreAuger:2013eja,YakutskHarmonics}. Therefore, our  results suggest that the average magnetic field  in voids must be $B < 0.1 \ \rm nG$ to reconcile with observations under the assumption that most UHECRs at $\sim$ EeV have an extragalactic origin. 
	
\section{Discussion and conclusions}
\label{sec:conclusions}
	In this work we investigated the role of cosmic magnetic fields and of the source distribution on observed UHECRs with energies $\geq 10^{18}$ eV, using cosmological MHD simulations. 
	We tested the outcome of several scenarios for the origin of extragalactic magnetic fields (e. g. primordial versus astrophysical) and their observable signatures in the properties of UHECRs, assuming a pure proton composition and an entirely extragalactic origin.
\\ \\
	As very important caveats for theoretical modelling,  we have found that simulations of UHECR propagation are prone to geometrical artefacts, i. e. the measured deflection of UHECRs from nearby sources is dominated by the finite size of simulated observers. This is at variance  with the interpretation by \citet{Ryu_deflection}, who suggested that nearby sources of UHECRs may be located within the same filament as the observer, based on their large deflection. 
\\ \\ \\
	We have also found that the arrival energy spectrum of UHECRs above $10^{18} \eV$ is independent of the underlying magnetic field model.
	Above the GZK-threshold, $E > 4 \cdot 10^{19} \eV$, the angular power spectrum of events do not show significant effects of the underlying magnetic field model, and the spectra of all models are consistent with the magnetic field free case.
	For average field strength in voids $B \geq 0.1 \ \rm nG$, we have found deviations from isotropy at a few EeV in the angular power of the lowest multipoles.
	Therefore, magnetic fields in filaments and outskirts of clusters do not affect the observable properties of UHECRs in any significant way, for the range of models considered here. 
	Furthermore,  we also find that cosmic voids only minimally affect the propagation of UHECRs above the GZK-threshold for EGMF models that are in agreement with the limits from the CMB analysis \citep{PLANCK2015}.
	On the other hand, we have found that the distribution of the most nearby sources within $50 \Mpc$ dominates the anisotropy signal.
\\
	The angular distance between pairs of events, as well as between events and the most nearby source,
are also dominated by the most nearby sources.
	We suggest that these properties can be used to correlate larger catalogs of observed sources with the observed distribution of UHECRs events.\\
	In this work, we did not account for the deflection of the magnetic field of the Milky Way.
	However, the arrival energy spectrum should not be affected by the galactic magnetic field at these energies because the travel time and hence the energy losses do not increase significantly.
	The angular power spectrum of a pure proton scenario was also shown to be mostly unaffected by the galactic magnetic field \citep{arXiv:1411.2486v2}.
	The separation angle plots should instead be more affected by galactic deflection, but only for the most nearby sources, which contribute most of the non-isotropic signal.
\\ \\
	In summary, we have found that for a pure proton composition, the cosmic magnetic fields in cosmic structures  have little to no influence on the observable properties of UHECRs. Instead,  the distribution of most nearby sources within $50 \Mpc$ is found to dominate the anisotropy of events, hence enabling "UHECRs astronomy".
	The use of tailored simulations to reproduce the real distribution of structures around the Milky Way might offer a chance to highlight any residual signature of extragalactic magnetic fields, by minimising the impact of cosmic variance. 
	We further found that the absence of anisotropy of UHECRs at a few EeV can be used to impose an upper limit on the average strength of primordial magnetic fields, of the order of $\sim 0.1 \ \rm nG$ comoving.
\\ \\
	All these results should also be tested with heavier composition because particles with a higher charge are deflected more strongly and could therefore be more sensitive to the different magnetic field models.
	It is further crucial to add a prediction of the deflection inside the galactic magnetic field of the observer to see how stable the results are against this further local deflection.
	We plan to continue our work using constrained models of the local universe, and  to probe different source distributions that coincide with catalogues of potential sources of UHECRs.
	
\section*{acknowledgments}

We would like to acknowledge the developers of CRPropa for making the code publicly available.
The cosmological simulations presented in this work were performed using the {\enzo} code (http://enzo-project.org), and were 
partially produced at Piz Daint (ETHZ-CSCS, http://www.cscs.ch) in the Chronos project ID ch2 and s585, and on the JURECA supercomputer at  the NIC of the Forschungszentrum J\"{u}lich,  under allocations no. 7006 and 9016 (FV) and 9059 (MB).  We also acknowledge the use of the Golem Cluster at Hamburger Sternwarte for analysis. FV acknowledges personal support from the grant VA 876/3-1 from the Deutsche Forschungsgemeinschaft (DFG). FV and MB also acknowledge partial support from
the grant FOR1254 from DFG.
This work was supported by the DFG through collaborative research centre SFB 676, by the Helmholtz Alliance for Astroparticle Physics (HAP) funded by the Initiative and Networking Fund of the Helmholtz Association and by the Bundesministerium f\"ur Bildung und Forschung (BMBF).
We thank R. Clay, R. Alves Batista and J. Devriendt  for useful comments on the manuscript.

\bibliographystyle{mn2e}
\bibliography{cites}

\label{lastpage}

\appendix

\end{document}